\documentclass[11pt]{article}
\usepackage{amsmath,amssymb,color,epsfig,cite}
\usepackage{graphicx}
\usepackage{subfigure}
\usepackage{setspace}

\usepackage{amsfonts}

\makeatletter
\@addtoreset{equation}{section}
\makeatother

\newcommand{\be}{\begin{equation}}
\newcommand{\ee}{\end{equation}}
\newcommand{\bea}{\setlength\arraycolsep{2pt} \begin{eqnarray}}
\newcommand{\eea}{\end{eqnarray}}
\newcommand{\nn}{\nonumber}

\def\ft#1#2{{\textstyle{\frac{\scriptstyle #1}{\scriptstyle #2} } }}
\def\fft#1#2{{\frac{#1}{#2}}}

\def\0{{\sst{(0)}}}
\def\1{{\sst{(1)}}}
\def\2{{\sst{(2)}}}
\def\3{{\sst{(3)}}}
\def\4{{\sst{(4)}}}
\def\5{{\sst{(5)}}}
\def\6{{\sst{(6)}}}
\def\7{{\sst{(7)}}}
\def\8{{\sst{(8)}}}
\def\9{{\sst{(9)}}}

\def\sst#1{{\scriptscriptstyle #1}}

\begin{document}



\begin{center}
{\large {\bf Holographic OPE Coefficients from AdS Black Holes with Matters}}

\vspace{10pt}
Yue-Zhou Li, Zhan-Feng Mai and H. L\"u

\vspace{15pt}

{\it Center for Joint Quantum Studies and Department of Physics,\\
School of Science, Tianjin University, Tianjin 300350, China}

\vspace{30pt}

\underline{ABSTRACT}
\end{center}

We study the OPE coefficients $c_{\Delta, J}$ for heavy-light scalar four-point functions, which can be obtained holographically from the two-point function of a light scalar of some non-integer conformal dimension $\Delta_L$ in an AdS black hole. We verify that the OPE coefficient $c_{d,0}=0$ for pure gravity black holes, consistent with the tracelessness of the holographic energy-momentum tensor.  We then study the OPE coefficients from black holes involving matter fields. We first consider general charged AdS black holes and we give some explicit low-lying examples of the OPE coefficients. We also obtain the recursion formula for the lowest-twist OPE coefficients with at most two current operators. For integer $\Delta_L$, although the OPE coefficients are not fully determined, we set up a framework to read off the coefficients $\gamma_{\Delta,J}$ of the $\log(z\bar{z})$ terms that are associated with the anomalous dimensions of the exchange operators and obtain a general formula for $\gamma_{\Delta,J}$. We then consider charged AdS black holes in gauged supergravity STU models in $D=5$ and $D=7$, and their higher-dimensional generalizations.  The scalar fields in the STU models are conformally massless, dual to light operators with $\Delta_L=d-2$. We derive the linear perturbation of such a scalar in the STU charged AdS black holes and obtain the explicit OPE coefficient $c_{d-2,0}$. Finally, we analyse the asymptotic properties of scalar hairy AdS black holes and show how $c_{d,0}$ can be nonzero with exchanging scalar operators in these backgrounds.

\vfill {\footnotesize liyuezhou@tju.edu.cn\ \ \ \ zhanfeng.mai@gmail.com\ \ \ \ mrhonglu@gmail.com}

\pagebreak

\tableofcontents
\addtocontents{toc}{\protect\setcounter{tocdepth}{2}}


\newpage
\section{Introduction}
\label{sec:intro}

The AdS/CFT correspondence establishes an insightful routine to investigate a strongly coupled conformal field theory (CFT) by using appropriate weakly coupled gravity in anti-de Sitter (AdS) spacetime and vice versa \cite{Maldacena:1997re}. Originally, the AdS/CFT correspondence is typically referred to as the duality between type IIB superstring in AdS$_5\times S^5$ and ${\cal N}=4$, $d=4$ super Yang-Mills theory. The holographic principle is expected to be more general and can apply to a variety of gravity theories even without supersymmetry, and indeed it has passed a large amount of tests at the AdS scale, i.e.~the locality holds at the scale that is never shorter than the AdS radius $\ell$ \cite{Heemskerk:2009pn}. The results include the correct structures of two-point functions, three-point functions \cite{Gubser:1998bc,Witten:1998qj}, conformal anomalies \cite{Henningson:1998gx,Henningson:1998ey} in CFTs that are fixed by the virtue of conformal symmetry. Typically, even though the structures are the same, different gravity theories may lead to different CFT data. Thus gravities can be served as effective CFTs. By finding relations and bounds from the holographic CFT data that follow exactly the same pattern regardless of the specific details of a gravity theory, some universal properties of CFTs can be revealed. Known examples include the controlling pattern of shear-viscosity/entropy ratio and entanglement entropy by central charges \cite{Kats:2007mq,Banerjee:2009fm,Hung:2011nu,Li:2019auk}, central charge relations \cite{Li:2018drw,Bueno:2018yzo,Lu:2019urr}.

Below the AdS scale where the higher-point correlation functions ($\geq4$) come out to be visible, the generality of AdS/CFT becomes highly nontrivial. Fortunately, it was argued that any large $N$ CFT with a parametrically large conformal dimensions for single-trace higher spin operator (spin $J>2$) can have a weakly coupled gravity dual \cite{Heemskerk:2009pn}. With this generality in mind, it is then natural to follow the same logic for the AdS-scale holography to find universal properties of CFTs by studying generic gravity theories.

The simplest case is the four-point functions. Typically, the four-point functions can be decomposed into conformal blocks which are completely determined by conformal symmetry with theory dependent OPE coefficients. (See \cite{Qualls:2015qjb,Rychkov:2016iqz,Simmons-Duffin:2016gjk} and also Appendix \ref{conformalblock} for a brief pedagogical review.) One may then expect to study the four-point functions from the bulk to recover the conformal blocks and read off the OPE coefficients, and investigate some possible universal pattern. However, although the holographic conformal blocks as the geodesic Witten diagram were studied extensively in literature (here is the incomplete list \cite{Hijano:2015zsa,Hijano:2015qja,Alkalaev:2015lca,Castro:2017hpx,Dyer:2017zef,Chen:2017yia,Kraus:2017ezw},) explicitly computing them for quite general classes of higher-derivative gravities is rather challenging. The issue can be greatly simplified by considering the special case of heavy-light four-point functions in the heavy limit \cite{Fitzpatrick:2019zqz}. In this case, the four-point function can be treated as the two-point function of the light operators under overwhelmingly heavy states that can be viewed as black hole backgrounds in the bulk. This special case avoids the difficult task for addressing the holographic conformal blocks of four-point functions directly. Focusing on the pure gravity black holes and deriving the holographic OPE coefficients, Ref.~\cite{Fitzpatrick:2019zqz} found that the lowest-twist OPE coefficients for multi-stress tensors are universal. We shall review this in section \ref{OPE}.

On the other hand, higher-point correlation functions, such as four-point functions of strongly coupled CFT can be studied further than the structures without referring to any specific theory by bootstrap program. (See \cite{Poland:2018epd} for a recent review.) The consistency conditions emphasized in bootstrap program, for example, the unitarity \cite{Mack:1975je,Minwalla:1997ka}, the crossing symmetry \cite{Rattazzi:2008pe,Caracciolo:2009bx,Komargodski:2012ek,Fitzpatrick:2012yx,Alday:2015ewa} and averaged null energy condition (ANEC) \cite{Hofman:2008ar,Belin:2019mnx,Kologlu:2019bco}, can universally constrain the spectrum and CFT data beyond two and three-point functions. These strong constraints in CFTs shall, inversely, be mapped to the constraints to the bulk theories to select those consistent (quantum) gravities with sensible CFT duals. For example, the crossing equation was used to restrict the interaction terms for bulk gravity theories \cite{Heemskerk:2009pn}. Of course, the simplest consistency condition in CFT should be that the stress tensor is traceless. This implies in particular that the OPE coefficient $c_{d,0}$ must vanish when the CFT does not have an additional scalar operator of conformal dimension $d$.  Indeed it was verified \cite{Fitzpatrick:2019zqz} that $c_{4,0}=0$ from pure gravity black holes.  In this paper shall verify the consistency for general $c_{d,0}$.

The main motivation of this paper is to study the holographic OPE coefficients for general black holes involving matter fields. We find that the patterns of lowest-twist OPE coefficients becomes more interesting, and the universality in CFT should be reconsidered. The paper is organized as follows.
\begin{itemize}

\item In section \ref{OPE}, we begin with a review of the proposal for the heavy limit of holographic heavy-light scalar four-point functions and the corresponding holographic OPE coefficients. We review the construction and conclusions of \cite{Fitzpatrick:2019zqz} for pure gravity backgrounds in some detail. Moreover, we verify the consistency that the OPE coefficient $c_{d,0}$ associated with the trace of stress tensor does vanish for general $d$. This fact motivates us to consider black holes with matters such that more primary operators can engage in, for instance, contributing to $c_{d,0}\neq0$.

\item In section \ref{OPEcharge}, we consider AdS black holes charged under a Maxwell field in a general class of high-derivative gravity-Maxwell theories. In addition to the stress tensor, we find that the conserved current operator can also appear in the conformal blocks. Some explicit low-lying examples in $d=4$ and $d=6$ are presented to gain insights. Moreover, we obtain a recursion formula for computing the OPE coefficients involving at most two currents in general even dimensions $d$. We find some clear patterns in the OPE coefficients with their dependence on $(f_0, \tilde{f}_0)$, the integration constants proportional to the black hole mass and charge respectively.  We conjecture that in general, the lowest-twist OPE coefficients should be $c_{\Delta=n_1 d + 2 n_2 (d-1), J=2n_1 + 2n_2} \propto f_0^{n_1} \tilde f_0^{n_2}$. This generalizes to the results of \cite{Fitzpatrick:2019zqz} where only the $n_2=0$ case was considered.

\item In section \ref{integerDelta}, we discuss the subtlety for integer $\Delta_L$. In this case there will be logarithmic terms appearing in the solutions of the bulk linearized equation associated with the light operator. We formulate the construction for dealing with the logarithmic terms and give a few examples. The logarithmic terms $\log(z\bar{z})$ can be naturally interpreted as anomalous dimensions. We find that although the OPE coefficients mixed with double-trace operators cannot be fully determined, the anomalous-dimension related coefficients $\gamma_{\Delta,J}$ can be. We exhibit and prove a formula of determining $\gamma_{\Delta,J}$ in terms of the residue of OPE coefficients for non-integer $\Delta_L$.

\item In section \ref{gauge-sg}, we turn to consider gauged supergravities and the supergravity inspired models where there are additional scalar fields involved in the black hole backgrounds. We study two cases: (1) the light operator is outside the supergravity, (2) the light operator is part of the supergravity theory. For both cases, we find that in $D=5$ gauged supergravity, even though $c_{4,0}\neq 0$, there is no inconsistency because the additional scalars involved in the black hole have conformal dimensions $\Delta=2$ and they can contribute to $c_{4,0}$. Furthermore, we find that for the case (1), although the spectrum has $\Delta=d-2$ operators, $c_{d-2,0}$ is nevertheless vanishing. This issue can be resolved, however, by considering case (2) where the light operator is within the supergravity theory. For those light operators lying in the supergravity, $\Delta_L$ is, inevitably, an integer, for which we exhibit explicit examples for $\gamma_{\Delta,J}$ and verify the equation found in section \ref{integerDelta} again.

\item In section \ref{scalarhair}, inspired by the study of supergravity cases in section \ref{gauge-sg}, we turn to consider the general scalar hairy black holes. We show that we can always have $c_{d,0}\neq 0$ by considering scalar hairy black holes which contain either the operators with $\Delta=d$ or the operators with $\Delta=d/2$.

\item In section \ref{conc}, we summarize the paper and present the outlook for future investigations.

\item In Appendix \ref{conformalblock}, the preliminary knowledge of conformal blocks is sketched.

\item In Appendix \ref{freescalar}, the linearized equation for scalars in the variables considered in this paper is presented.
\end{itemize}

\section{OPE coefficients from holography}
\label{OPE}

In this section, we study the formalism using holographic technique to compute the heavy-light four-point functions in the heavy limit.  The formalism was developed in \cite{Fitzpatrick:2019zqz} for the case involving two light scalar operators and two heavy operators that are dual to the AdS planar black holes constructed in the pure gravity sector. We begin with the review of the formalism and then examine the consistency of the vanishing $c_{d,0}$ for general even $d$.

\subsection{Four-point functions and conformal blocks}

We consider heavy-light four-point functions that contain two heavy operators $\mathcal{O}_H$ with parametrically large conformal dimensions $\Delta_H\sim C_T$, where $C_T$ is the overall coefficient associated with the two-point function of the stress tensor $T_{\mu\nu}$. Two light operators $\mathcal{O}_L$ have much smaller conformal dimensions, namely $\Delta_L\ll C_T$. The four-point functions can be decomposed into conformal blocks $G$. In appendix \ref{conformalblock}, we give a short review on conformal blocks and their properties associated with scalar four-point functions. In $s$-channel in the conformal frame, defined by (\ref{cf}), the four-point function can be decomposed as
\bea
\langle \mathcal{O}_H \mathcal{O}_L \mathcal{O}_L \mathcal{O}_H\rangle=(z\bar{z})^{-\fft
{\Delta_H+\Delta_L}{2}}\sum_{\mathcal{O}}c_{\Delta,J}\,G^{\Delta_{HL},-\Delta_{HL}}_{\Delta,J}(z,\bar{z})
\,,\label{heavy-light-4pt}
\eea
where $\Delta_{HL}=\Delta_H-\Delta_L$, $z$ and $\bar{z}$ are related to cross-ratios and $c_{\Delta,J}$'s are the products of two OPE coefficients, now commonly referred to simply as OPE coefficients. Note that $c_{\Delta, J}$'s also depend on $\Delta_L$ and $\Delta_H$.

In the holographic picture, the state excited by a heavy operator in the heavy limit can be viewed as some asymptotically AdS spacetime while a light operator is some perturbation in this background. The most important backgrounds are perhaps the AdS black holes which can be therefore interpreted as excited states $|{\rm BH}\rangle\simeq\mathcal{O}_H |0\rangle$. Although the formalism we are going to discuss involves only the asymptotic structures and hence its application is not limited to black hole geometries.  Nevertheless we refer to all asymptotic AdS geometries as black holes for simplicity. Thus holographically we can treat the four-point function (\ref{heavy-light-4pt}) as the two-point function in the black hole state, namely
\be
\langle \mathcal{O}_H \mathcal{O}_L \mathcal{O}_L \mathcal{O}_H\rangle\simeq \langle  \mathcal{O}_L \mathcal{O}_L \rangle_{\rm BH}\,.
\ee
In other words, the problem reduces to compute the linear perturbation of the corresponding dual bulk field in the black hole background and derive the two-point function using the standard holographic dictionary.

It turns out that it is advantageous to compute the four-point function in t-channel instead, namely
\be
\langle  \mathcal{O}_L \mathcal{O}_L \rangle_{\rm BH}=((1-z)(1-\bar{z}))^{-\Delta_L}\sum_{\mathcal{\tilde{O}}}\tilde{c}_{\Delta,J}
\,G^{0,0}_{\Delta,J}(1-z,1-\bar{z})\,,\label{t-channel1}
\ee
which should be the same as the $s$-channel result because of crossing symmetry \cite{Simmons-Duffin:2016gjk}.  The four-point function in the light-cone limit $z\rightarrow 1$ acquires a simplification since the heavy exchanged operators in OPEs in the t-channel would not survive. It follows that the exchanged operators $\tilde {\cal O}$ in $t$-channel are necessarily light operators with $\Delta\sim \Delta_L$ in the spectrum.

When there is no confusion, for convenience, we simply drop off the tilde of the OPE coefficient $\tilde c$ in (\ref{t-channel1}) and replace $z$ by $1-z$ such that the light-cone limit becomes $z\rightarrow 0$, i.e.
\be
\langle  \mathcal{O}_L \mathcal{O}_L \rangle_{\rm BH}=(z\bar{z})^{-\Delta_L}\sum_{\mathcal{O}}c_{\Delta,J}\,G^{0,0}_{\Delta,J}(z,\bar{z})\,.\label{t-channel2}
\ee
The fact that only $G^{0,0}_{\Delta, J}$ appears in the decomposition is sufficient to indicate that it is the $t$-channel four-point function. (For more relevant properties of conformal blocks $G^{0,0}_{\Delta,J}(z,\bar{z})$, see Appendix \ref{conformalblock}.) Thus the holographic technique now amounts to calculating the two-point functions in the black hole backgrounds, comparing with the definition of the conformal blocks (\ref{t-channel2}), and reading off the OPE coefficients.

This was carried in \cite{Fitzpatrick:2019zqz} for a free massive scalar in AdS planar black holes constructed in the pure gravity sector for even $d$ dimensions. It was found that the lowest-twist OPE coefficients, i.e. $c_{\Delta,J}$ with the minimum twist $\tau=\Delta-J$, are somehow universal in the sense that they do not depend on the details of the gravity theory under consideration. Since many of the results will be useful for the rest of the paper, we shall give a detail review of the construction in subsection \ref{const-G} and \ref{conc-G}.

The reason that one can treat the static $|{\rm BH}\rangle$ as the  dual to a (heavy) scalar operator that appears in the four-point function (\ref{4pt}) is that it is specified by the mass only with no spin.  As we shall elaborate in subsection \ref{comments}, there is a consistency check that the holographic OPE coefficient $c_{d,0}$ must vanish for black holes in the pure gravity sector. This was shown the case for $d=4$ in \cite{Fitzpatrick:2019zqz}.  We shall prove that $c_{d,0}=0$ for general even dimensions in subsection \ref{comments}, before we study more general matter supported black holes.

\subsection{The construction in pure gravity backgrounds}
\label{const-G}

As in \cite{Fitzpatrick:2019zqz}, we consider here gravity minimally coupled to a free massive scalar
\be
{\cal L}=\sqrt{|g|} \Big(R-2\Lambda+L(R_{\mu\nu\rho\sigma})-\ft{1}{2}(\partial\phi)^2-m^2 \phi^2\Big)\,,\qquad \Lambda=\fft{d(d-1)}{2\ell_0^2}\,,\label{gravity-nomatter}
\ee
where $L(R_{\mu\nu\rho\sigma})$ represents the generic higher-order curvature polynomials, and $\ell_0$ is the bare AdS radius. For appropriate $L$, the theory admits an AdS vacuum of certain radius $\ell$.  In the Euclidean signature and in planar coordinates, it is given by
\be
ds^2 = \fft{dr^2}{r^2} + r^2 (dt^2 + du^2 + u^2 d\Omega_{d-2}^2)\,.\label{adsvac}
\ee
Here for simplicity, we set the AdS radius to unit. The boundary metric is assumed to be spherically symmetric.
It is useful to introduce complex light cone coordinates $(z,\bar z)$:
\be
t=-\fft12 (z + \bar z)\,,\qquad u=\fft{\rm i}2 (z-\bar z)\,.\label{tutoz}
\ee
These are precisely related to the cross ratios in the conformal frame (\ref{cf}) discussed in Appendix \ref{conformalblock}.

Maintaining the spherical symmetry in (\ref{adsvac}), one can construct Euclidean AdS planar black holes (with $\phi$ remaining zero):
\be
ds^2=r^2f(r)dt^2+\fft{1}{r^2h(r)}dr^2+r^2 (du^2 + u^2 d\Omega_{d-2}^2)\,.\label{metric}
\ee
For Einstein gravity, we have $h=f=1-f_0/r^d$, namely the Schwarzschild-AdS planar black hole. For general ${\cal L}(R_{\mu\nu\rho\sigma})$, the linear spectrum in the AdS vacuum contains a massive scalar mode and a ghost-like massive spin-2 mode, in addition to the usual massless graviton. The asymptotic behavior of the function $h$ and $f$ can be very complicated and difficult to classify when the massive modes are turned on.
In massless gravities, where the massive modes are decoupled \cite{Li:2018drw,Li:2019auk}, we have in general
\be
f(r)=1-\fft{f_0}{r^d}-\fft{f_d}{r^{2d}}-\cdots\,,\qquad h(r)=1-\fft{h_0}{r^d}-\fft{h_d}{r^{2d}}-\cdots\,.
\label{hfexpansion}
\ee
To be precise we must have $f_0=h_0$, but here we leave them independent for a more general discussion.  In quasi-topological gravities \cite{Oliva:2010eb,Myers:2010jv,Myers:2010ru,Li:2017ncu,Peng:2018vbe} or Einsteinian cubic gravities \cite{Bueno:2016xff,Hennigar:2016gkm,Bueno:2016lrh,Bueno:2018xqc} (only in $D=4$), one has $h=f$ by construction.

For the black holes in massless gravities, the parameter $f_0$ is related to the black hole mass. In the boundary CFT, $f_0$ has a universal interpretation in the sense that it only depends on the ratio $\Delta_H/C_T$, namely \cite{Kulaxizi:2018dxo,Karlsson:2019qfi,Fitzpatrick:2019zqz}
\be
f_0=\fft{4\Gamma(d+2)}{(d-1)^2 \Gamma(\fft{d}{2})^2}\fft{\Delta_H}{C_T}\,.\label{f0}
\ee
The equation of motion for the free scalar $\phi$ around the black hole is given by
\be
(\Box-m^2)\phi=0\,,\qquad m^2=\Delta_L(\Delta_L-d)\ge m_{\rm BF}^2=-\ft14 d^2\,,\label{phi1}
\ee
where we assume that $\Delta_L\ge \Delta_L - d$ so that $(\Delta_L-d, \Delta_L$) are the conformal dimensions associated with the source and response modes respectively.  The minimum conformal dimension is thus $\Delta_L=\ft12 d$, corresponding to saturating the Breitenlohner-Freedman (BF) mass bound $m_{\rm BF}^2$ for the scalar $\phi$.

According to the standard AdS/CFT dictionary, the solution of (\ref{phi1}) in the background (\ref{metric}) gives rise to the bulk-to-boundary propagator $\Phi(r,t,u)$ in which the coefficient of $1/r^{\Delta_L}$ is the two-point function. The coordinates $(t,u)$ is related to $(z,\bar{z})$ in the conformal frame (\ref{cf}) by (\ref{tutoz}). As we can see in Appendix \ref{conformalblock}, the expression for conformal blocks can be complicated and few can be expressed analytically in closed forms. For general situations, one typically considers the OPE limit, namely taking $z\sim \bar z\rightarrow 0$.  In this limit, the conformal blocks can be given order by order, e.g.~(\ref{G00}) for $G^{00}_{\Delta,J}$.  The same situation arises for the bulk perturbation and one can solve the linear equation in the OPE limit.  To do so, Ref.~\cite{Fitzpatrick:2019zqz} made a change of coordinates
\be
w^2=1+r^2 (t^2+u^2)=1+r^2 z\bar{z}\,,\qquad \hat{u}=ru =\fft{\rm i}{2} r(z-\bar z)\,.\label{change-va}
\ee
In this coordinate system, the solution  to (\ref{phi1}) in the AdS vacuum (\ref{adsvac}) can be expressed simply as
\be
\Phi_{\rm AdS}=(\fft{r}{w^2})^{\Delta_L}\sim \fft{(z\bar z)^{-\Delta_L}}{r^{\Delta_L}} + \cdots\,,\qquad \hbox{for}\qquad r\rightarrow \infty\,.
\ee
Thus in the pure AdS background, the light scalar propagator is simply $(z\bar{z})^{-\Delta_L}$. Comparing to (\ref{t-channel2}), it is natural to factorize the bulk-to-boundary propagator in general asymptotic AdS backgrounds as
\be
\Phi(r,w,\hat{u})=\Phi_{\rm AdS}G(r,w,\hat{u})\,.\label{factorize}
\ee
Then the function $G(r,w,\hat u)$ in the $r\rightarrow \infty$ limit is precisely the conformal block.  In other words, the holographic dictionary now reduces to \cite{Fitzpatrick:2019zqz}
\be
\sum_{\mathcal{O}}c_{\Delta,J}\,G^{0,0}_{\Delta,J}(z,\bar{z}) = \lim_{r\rightarrow \infty} G(r,w,\hat u)\,.
\label{holodict}
\ee
Note that the right-hand side of the above is convergent and therefore the subleading terms in $\Phi_{\rm AdS}$ does not contribute to the conformal block. The linear equation for $G(r,w,\hat u)$ is given in (\ref{eqG}) in Appendix \ref{freescalar}. Recall the conformal block series expansion (\ref{G00}), it then follows from (\ref{change-va}) that the conformal block with a certain $\Delta$ is closely attached to the coefficient of $1/r^{\Delta}$ in $G(r,w,\hat{u})$.

For the general theories (\ref{gravity-nomatter}) but restricted to massless gravities, there are two sets of operators that can exchange in the OPE expansions in the scalar four-point function. The first set is multi-stress tensor operators $T^n$, denoting all possible multiplications of the stress tensor (of conformal dimension $d$) to the $n$'th order. They contribute to the conformal blocks with $\Delta=n d$. The second set is the double-trace operators $[\mathcal{O}_L]^\Delta_J$ with spin $J$ and conformal dimension $\Delta=2\Delta_L+2n+J$ composed by $\mathcal{O}_L$
\be
[\mathcal{O}_L]^\Delta_J=\mathcal{O}_L \Box^n\partial_{\mu_1}\cdots\partial_{\mu_J}\mathcal{O}_L\,.\label{double}
\ee
The near boundary expansion for $G(r,w,\hat{u})$ therefore should take the form
\bea
 G(r,w,\hat{u})=1&+&G^T(r,w,\hat{u})+G^{L}(r,w,\hat{u})\,,\nn\\
G^T(r,w,\hat{u})=\fft{1}{r^d}\sum_{i\in\mathbb{N}}\fft{G^T_i(w,\hat{u})}{r^{id}}\,,&&  G^{L}(r,w,\hat{u})=\big(\fft{w}{r}\big)^{2\Delta_L}
\sum_{i\in 2\mathbb{N}}\fft{G^L_i(w,\hat{u})}{r^i}\,,\label{ansatz1}
\eea
where $1$ represents the identity block\footnote{Note that we are working on the $t$-channel here, in which the identity block is actually contributed by an infinite number of operators in the $s$-channel.}. Note that both $w$ and $\hat u$ depend on $r$, it follows that $G^{(T,L)}(r,w,\hat u)$ are both non-vanishing in the $r\rightarrow \infty$ limit.  When $\Delta_L$ is not an integer, the two sets are independent. As in \cite{Fitzpatrick:2019zqz}, we shall focus on the the case of non-integer $\Delta_L$. We shall comment on the case of integer $\Delta_L$ later.

As mentioned above, $G^T_i(w,\hat{u})$'s directly relate to conformal blocks with $\Delta=(1+i)d$ and $G^L_i(w,\hat{u})$ relate to those with $\Delta=2\Delta_L+i=2\Delta_L+2n+J$. Since the conformal block coefficients are non-zero only for even spin $J$, it follows that $i$ for $G^T_i$ must be even numbers.  Since $G^T_i$ and $G^L_i$ should be related to the conformal blocks with certain $\Delta$, they must take the polynomials of $\hat{u}$,
\bea
G^T_i=\sum_{j\in 2\mathbb{N}}^{2(1+i)} a_{ij}(w)\hat{u}^j\,,\qquad G^L_i=\sum_{j\in 2\mathbb{N}}^{i}b_{ij}(w)\hat{u}^j\,.\label{ansatz2}
\eea
The truncation to the finite orders of the polynomials of $\hat{u}$ in (\ref{ansatz2}) is subtle. To see this, one notices that the relevant term giving conformal block with $\Delta$ is $w^{\Delta-m}\hat{u}^m$. If there were no such truncations, we would have
\be
\fft{1}{r^{\Delta}}\sum_{m=-\infty}^{\infty}w^{\Delta-m}\hat{u}^m\sim \sum_{m=-\infty}^{\infty}(z\bar{z})^{\fft{\Delta-m}{2}}(z-\bar{z})^m\,.\label{trun}
\ee
However, as can be seen from (\ref{powerlaw}), the lowest power for $z$ in conformal blocks should be $\fft12 (\Delta-J)$, we then have $m\leq J$. Thus, for the multi-stress set $T^n$, we have the $m\leq2(1+i)$ truncation and for the double-trace set (\ref{double}) we have the truncation that the $m\leq J=i-2n\leq i$ and that the coefficients of the higher-$n$ terms vanish.

We can now substitute (\ref{ansatz1}) and (\ref{ansatz2}) into the scalar equation (\ref{eqG}) and solve for $a_{ij}$ and $b_{ij}$. Taking the $r\rightarrow\infty$ limit for the solution $G$ and then comparing the result to the conformal blocks (\ref{G00}), we can read off the OPE coefficients in terms of $(f_i,h_i)$ in (\ref{hfexpansion}). In practice, however, the scalar equation (\ref{eqG}) can only determine $a_{ij}(w)$ while it has no restriction on $b_{ij}(w)$. As we shall see in further examples, this may related to the fact that $\phi$, associated with the light operator, does not involve in the black hole background. In fact, as we explain in Appendix \ref{freescalar}, the equation for $G^T$ contain a source supplemented by the background metric whilst the $G^L$ function remains source free and hence cannot be determine.  The absence of any source is related to the fact that $\phi$ does not involve in the construction the background metric and hence there is no falloffs of the type $1/r^{\Delta_L}$ in the metric.

Consequently, the construction based solely on the asymptotic structure can only reveal the holographic OPE coefficients for multi-stress tensor contributions of heavy-light four-point functions in the heavy limit, while the double-trace contributions are far from clear. For this reason, in this paper we in general simply drop the $G^L$ terms altogether (when it is source free), except in a few special cases where $G^L$ terms cannot be avoided.

It turns out that in even $d$ dimensions, $a_{ij}(w)$ can be polynomials of $w$ \cite{Fitzpatrick:2019zqz}
\be
a_{ij}(w)=\sum_{k=-2(1+i)}^{(1+i)d-j} a_{ijk}w^k\,,\label{ansatz3}
\ee
where the lower bound of the polynomial truncation for $w$ is the same but in opposite sign to the upper bound of the truncation for $\hat{u}$. This is because we simply let $\Delta-m\rightarrow m$ in (\ref{trun}), and then we have $m\geq \Delta-J\geq -J$. Throughout this paper, we shall consider only even $d$ dimensions such that we have the manageable polynomial ansatz (\ref{ansatz3}).  Consequently the exchanged multi-stress tensor operators all have even $\Delta=n d$ conformal dimensions.

In fact, the solution $a_{ijk}$ contain poles $\Delta_L-n$ where $n$ belongs to some finite set of nature numbers. Thus when $\Delta_L$ is itself an integer, the stress-tensor part of the contributions diverges. The requirement that the full solution be analytic in $\Delta_L$ indicate that $b_{ij}$ should also contain the same poles and all poles shall cancel each other such that the full solution is smooth. We leave further discussions on this issue in section \ref{integerDelta}.

\subsection{Lowest-twist OPE from pure gravities}
\label{conc-G}

The important conclusion in \cite{Fitzpatrick:2019zqz} is that for the pure gravity AdS black holes, it turns out that the lowest-twist OPE coefficients of multi-stress tensor operators are universal with only the dependence of $f_0$. For a given set of product operators of conformal dimensions $\Delta$, the lowest-twist operator has the maximum possible $J$ such that the twist $\tau=\Delta-J$ is minimum. For the multi-stress tensors $T^n$ we consider, the conformal dimension is $\Delta=n d$ and the maximum spin is $J=2n$.  Thus the lowest twist operator is
\be
T_{(\mu_1\nu_1} \cdots T_{\mu_{n}\,\nu_{n})}:\qquad \Delta=nd\,,\qquad J=2n\,,\qquad\tau=n(d-2)\,.
\ee
To isolate the lowest-twist contributions, we recall the analysis right below (\ref{trun}). The highest power of $\hat{u}$ is the highest spin $J$ for each conformal dimension $\Delta$; therefore, it is advantageous to introduce $\xi$ by $\hat u=r^{d/2} \xi$.  In the large $r$ limit, then only the lowest-twist contributions become relevant at the leading order while all other contributions are suppressed. In other words, the ansatz of $G$ in this limit becomes
\be
G(r,w,\xi)=Q(w,\xi)+\mathcal{O}(\fft{1}{r})\,,
\ee
and the scalar equation (\ref{eqG}) is reduced to be
\bea
&&16f_0 \xi^2 \Delta_L(\Delta_L+1) Q + \xi w^2 \big((d-2)(d+2-4\Delta_L) w^2 + 8 d\Delta_L\big) \partial_\xi Q\cr
&&-(d-2)^2 \xi^2 w^4 \partial_\xi^2 Q-4w\Big(w^2 \left(w^2 (d+1-2 \Delta_L)+2 \Delta_L -1\right) +
f_0\xi^2(1 + 4\Delta_L)\Big)\partial_w Q\cr
&&-4\xi w^3 (d + (2-d)w^2) \partial_\xi\partial_w Q +4w^2 (f_0 \xi^2 + w^2 - w^4)\partial_w^2 Q=0\,.
\label{reduced-eq-nomatter}
\eea
It is now straightforward to see that only $f_0$ of the bulk background enters the equation. The lowest-twist OPE coefficient depends only on $f_0$, which is universally proportional to the ratio $\Delta_H/C_T$ of the CFT parameters, as in (\ref{f0}).  The ansatz for $Q(w,\xi)$, following the analysis below (\ref{ansatz3}), is given by
\be
Q(w,\xi)=\sum_{m=-n}^{\fft{d-2}{2}n}\sum_{n\in \mathbb{N}}^{\infty}a_{nm}\xi^{2n}w^{2m}\,.\label{ansatz-redu}
\ee
Substituting it into (\ref{reduced-eq-nomatter}) yields a recursion relation for $a_{nm}$
\bea
a_{nm} &=& \fft{1}{4(m-nd)}\Big(\fft{(2(m+n-1)-nd)(2(m+n-1-\Delta_L)+(1-n)d)}{m-\Delta_L}a_{n,m-1}
\cr &&
\cr && -4(1-\Delta_L+m)f_0\, a_{n-1,m+1}\Big)\,,\qquad a_{0m}=\delta_{0m}\,.\label{recur-nomatter}
\eea
The OPE coefficients are related to $a_{nm}$ via
\be
c_{\Delta=nd,J=2n}=\fft1{2^{J}}\,a_{n,\fft{d-2}{2}n}\,.
\ee
As an example, we consider one stress-tensor $n=1$. Its (maximal) spin is $J=2$ and the lowest-twist OPE coefficient is \cite{Fitzpatrick:2019zqz}.
\bea
c_{d,2}=\fft{\Delta_L}{d+2}\fft{\Gamma(2+\fft{d}{2})^2}{\Gamma(3+d)} f_0=\fft{d^2\Delta_L \Delta_H}{4(d-1)^2C_T}\,.\label{cd2}
\eea

 We now would like to comment on the fact that the coefficients $b_{ij}$ cannot be determined by the equations of motion. This is not surprising since we are only looking at the solutions at the asymptotic expansion, without submitting them to the regularity constraints in the middle of the spacetime. What is highly non-trivial in the above approach is that the coefficients $a_{ij}$ can be nevertheless fully determined and hence all the OPE coefficients associated with the exchange of multi-stress tensor operators can be fully derived in even $d$ dimensions. This is the consequence of the coordinate choice $(w,\hat u)$ and the solution ansatz proposed by \cite{Fitzpatrick:2019zqz}.  As we can see in Appendix \ref{freescalar}, although the equation for $G$ is homogeneous without a source, the effect of the ansatz is that the equation for $G^T$ has a source depending on the metric functions while $G^L$ remains source free.

\subsection{Consistency of $c_{d,0}=0$}
\label{comments}

The OPE coefficient $c_{d,0}$ describes the exchange of a spin-0 operator of conformal dimension $d$. For pure (massless) gravity AdS black holes, together with a free scalar of non-integer $\Delta_L$, the only candidate is the trace of the energy-momentum tensor.  Thus we must have $c_{d,0}=0$.  This was shown for $d=4$ in \cite{Fitzpatrick:2019zqz}.  In this subsection, we examine the consistency for general even $d$.

First we examine the $d=4$ case in some detail. For AdS planar black holes constructed by pure massless gravities, we must take $f_0=h_0$. We can nevertheless pretend that they are different for generality, in which case, we have
\be
c_{4,0}=\fft{(f_0-h_0)(\Delta_L-4)\Delta_L}{120(\Delta_L-2)}\,.
\ee
On the other hand, the only possible operator contributing to $c_{4,0}$ in this setup is the trace of stress tensor $T_\mu^\mu$ which must vanish due to the conformal symmetry. Thus the condition $f_0=h_0$ which is always true for black holes constructed in pure massless gravities preserves the consistency $T_\mu^\mu=0$. This demonstrates that in $d=4$ AdS black holes constructed in the purely gravity sector is indeed dual to a scalar heavy operator in the heavy limit.  The conclusion above is in fact true for all even $d\ge 4$. To show this, we note that for $n=1$, with maximum $J=2$, $G(r,w,\hat u)$ involves at most quadratic $\hat u$, namely
\be
G(r,w,\hat{u})=1+\fft{1}{r^d} \Big(\sum_{k=-2}^{\fft{d}{2}} a_{k}w^{2k}+\sum_{k=-2}^{\fft{d-2}{2}}b_k\hat{u}^2w^{2k}\Big)\,.\label{G-d>4}
\ee
Substituting (\ref{G-d>4}) into equation (\ref{eqG}), the constant coefficients $a_n$ and $b_n$ can be solved exactly in arbitrary even $d$ dimensions. The $\hat{u}^2$-order gives
\be
b_{-1} = -\fft{f_0\Delta_L}{d+1}\,,\qquad
b_k = \fft{(d-2k)}{2(d-k)}b_{k-1}\,,\qquad k=2,3,\ldots\,.
\ee
For even $d$, the series terminates at $k=d/2$ and hence we have
\bea
b_k=-\fft{(d-2)f_0\Delta_L (2-\fft{d}{2})_{k-1}}{4(d^2-1)(2-d)_{k-1}}\,,\qquad-1\le k\le \ft12 d-1\,,\label{bn}
\eea
where $(i)_j$ is the Pochhammer polynomial
\be
(i)_j=\fft{\Gamma(i+j)}{\Gamma(i)}\,.
\ee
The $\hat{u}^0$-order terms give rise to the recursion relation for $a_n$:
\bea
a_{-1} &=&-\fft{(f_0+h_0)\Delta_L}{d+1}\,,\qquad a_0=\fft{(d-1)(f_0-h_0)+2(f_0+h_0)\Delta_L}{4(d+1)}\,,
\cr
a_1 &=& \fft{(f_0+h_0(d-\Delta_L)+df_0(\Delta_L-2))\Delta_L}{4(d^2-1)(\Delta_L-1)}\,,\cr
a_k &=&\fft{(2+d-2k)(k-\Delta_L-1)a_{k-1}-(d-1)b_{k-1}}{2(d-k)(k-\Delta_L)}\,,\qquad k\ge 2\,.\label{recur-a-d>4}
\eea
We thus end up with
\be
a_k=\fft{\Delta_L(-dh_0-f_0k+df_0(k+1-\Delta_L)+h_0\Delta_L)
(1-\fft{d}{2})_{k-1}}{4(d^2-1)(k-\Delta_L)(2-d)_{k-1}}
\,,\qquad k\geq-1\,.\label{an}
\ee
With the solution (\ref{an}) and (\ref{bn}), both OPE coefficients $c_{d,2}$ and $c_{d,0}$ can be read off. The $c_{d,2}$ result in (\ref{cd2}) can be reproduced precisely. We find that the coefficient $c_{d,0}$ is
\be
c_{d,0}=\frac{\sqrt{\pi } 2^{-d-1} \Delta_L  (d-\Delta_L ) \Gamma \left(\frac{d}{2}\right) (f_0-h_0)}{(d-2 \Delta_L ) \Gamma \left(\frac{d+3}{2}\right)}\,.\label{cd0-1}
\ee
It is then clear that whenever $f_0=h_0$, we have $c_{d,0}=0$, which signals the consistency for the construction. For a black hole constructed by pure gravity with only the massless graviton mode, we must have $f_0=h_0$. Furthermore, there is no more operator in addition to $T_\mu^\mu$ that has $(\Delta,J)=(d,0)$. Thus $c_{d,0}=0$ faithfully reflects that $T_\mu^\mu$ is vanishing for CFTs in flat spacetime. On the other hand, for black holes involving additional matter, it is not uncommon that $h_0\ne f_0$, in which case $c_{d,0}$ becomes non-vanishing.  It is of interest to examine that the corresponding exchange operator indeed has $(\Delta,J)=(d,0)$.

\section{OPEs from charged AdS black holes}
\label{OPEcharge}

\subsection{The construction and explicit examples}

In this section, we consider a general class of AdS black holes that are charged under a Maxwell field.
We consider a general class of theories of the following form
\be
L=R-2\Lambda-\ft{1}{4}F^2+\mathcal{L}(R_{\mu\nu\rho\sigma},F_{\mu\nu})-\ft{1}{2}(\partial\phi^2+m^2 \phi^2)\,,\label{gravity-maxwell}
\ee
where $\mathcal{L}(R_{\mu\nu\rho\sigma},F_{\mu\nu})$ represents the higher-order invariant polynomials of the curvature tensor and the strength $F_{\mu\nu}$ and hence matter and gravity can be generally non-minimally coupled. Higher-order gravity theories with higher-order Maxwell fields were studied extensively in the holographic context, see, e.g. \cite{Myers:2009ij,Feng:2015sbw}. As in the previous case, the black hole background associated with $|{\rm BH}\rangle$ does not involve $\phi$, the free scalar that is dual to the light operator ${\cal O}_L$. As in section \ref{const-G}, the (massless) gravitational sector gives rise to the leading falloff  $1/r^d$ and its integer powers in the metric functions $f$ and $h$ due to the dimension analysis. Now by including the Maxwell fields,  the black hole has additional falloffs $1/r^{2(d-1)}$ and its integer powers in $f$ and $h$. Furthermore, the dimension analysis implies additional terms $1/r^{nd+2m(d-1)}$ with positive integers $(n,m)$ are allowed. Thus charged AdS planar black holes have the following asymptotic expansion structure
\bea
f=1-\fft{f_0}{r^d}+\fft{\tilde{f}_0}{r^{2(d-1)}}-\fft{f_d}{r^{2d}}+\cdots\,,\qquad h=1-\fft{h_0}{r^d}+\fft{\tilde{h}_0}{r^{2(d-1)}}-\fft{h_d}{r^{2d}}+\cdots\,,\label{f-h-max}
\eea
where $\tilde{f}_0$ is proportional to $Q^2$ (the charge squared) of black holes. However, there is not yet any CFT interpretation analogous to (\ref{f0}) for $\tilde{f}_0$, and it is not supposed to be viewed as a universal CFT parameter. Note in our notation, we would like to denote all the new terms created by the existence of Maxwell fields with positive sign, e.g.  $+\tilde{f}_0$. In general, when the linear spectrum of the AdS background contains only the graviton and massless vector modes, we must have $h_0=f_0$ and $\tilde h_0 = \tilde f_0$. For now, we leave them different so that the results are applicable in the more general situation.

The scalar $\phi$ equation in the black hole background has the same form (\ref{phi1}), but now due to additional power appearing in (\ref{f-h-max}), the ansatz for $G(w,\hat{u})$ (\ref{ansatz1}) should involve new power terms of $1/r$. To be precise, we now have the power series
\be
G(r,w,\hat{u})=G^s(r,w,\hat{u})=\sum_{i,j\in\mathbb{N}}\fft{G_{ij}(w,\hat{u})}{r^{id+2j(d-1)}}\,,\qquad G_{00}=1\,,\label{ansatz1-maxwell}
\ee
where we simply drop the scalar double-trace mode contribution and denote the contributions from stress-tensor and conserved current as the ``short'' set $G^s$ (with identity block $G^{00}=1$).  The additional power laws with $1/r^{2(d-1)}$ in (\ref{ansatz1-maxwell}) indicate that by including Maxwell fields, the conserved current operator $\mathcal{J}$ with conformal dimension $\Delta=d-1$ and spin $J=1$ should also appear to exchange in the scattering process and thus be involved in conformal blocks. However, $\mathcal{J}$ can only appear in pairs due to the even spin requirement for the conformal blocks, it follows that the minimum $\Delta$ for the OPE coefficients that involve the Maxwell field is $2(d-1)$, which is again an even integer. Following the same procedure outlined in section \ref{OPE}, we find that the OPE coefficients can be derived. We now present some explicit low-lying examples in $d=4$ and $d=6$ for general non-integer $\Delta_L$.

\subsubsection{$d=4$}

In $d=4$, the near-boundary asymptotic expansions up to $1/r^{10}$ for $h$ and $f$ take the forms
\bea
f=1-\fft{f_0}{r^4}+\fft{\tilde{f}_0}{r^6}-\fft{f_4}{r^8}+\fft{f_6}{r^{10}}+\cdots\,,\qquad h=1-\fft{h_0}{r^4}+\fft{\tilde{h}_0}{r^6}-\fft{h_4}{r^8}+\fft{h_6}{r^{10}}+\cdots\,.\label{expansion-max-4d}
\eea
The structures dictates the ansatz for $G(r,w,\hat u)$:
\bea
&& G(r,w,\hat{u})=1+\fft{G_{10}(w,\hat{u})}{r^4}+\fft{G_{01}(w,\hat{u})}{r^6}+\fft{G_{20}(w,\hat{u})}{r^8}+\fft{G_{11}(w,\hat{u})}{r^{10}}+\cdots\,,
\cr &&
\cr && G_{10}=\sum_{j=-2}^{4-i}\sum_{i=0}^2 \alpha^{10}_{ij}\hat{u}^i w^j\,,\qquad G_{10}=\sum_{j=-2}^{6-i}\sum_{i=0}^2 \alpha^{01}_{ij}\hat{u}^i w^j\,,
\cr &&
\cr &&G_{20}=\sum_{j=-4}^{8-i}\sum_{i=0}^4 \alpha^{20}_{ij}\hat{u}^i w^j\,,\qquad G_{11}=\sum_{j=-4}^{10-i}\sum_{i=0}^4 \alpha^{11}_{ij}\hat{u}^i w^j\,.\label{d=4-max}
\eea
Substituting (\ref{d=4-max}) into equation (\ref{eqG}) will yield the solutions for all $(\alpha^{10}_{ij},\alpha^{01}_{ij},\alpha^{20}_{ij},\alpha^{11}_{ij})$. The results are too large to present here and we shall give only the OPE coefficients here. The simplest case is $\Delta=4$ and we have
\bea
c_{4,0}=\fft{(f_0-h_0)(\Delta_L-4)\Delta_L}{120(\Delta_L-2)}\,,\qquad c_{4,2}=\fft{f_0\Delta_L}{120}\,.
\eea
This is exactly the same as obtained in section 2. Thus for $\Delta=4$, including the Maxwell field in the bulk solution gives no contribution to the OPE coefficients for $\Delta=4$, and $c_{4,0}=0$ since we have $f_0=h_0$. This should be expected since the minimum $\Delta$ for the Maxwell field in the conformal block is 6.

We obtain explicit OPE coefficients and the corresponding exchanged operators $\Delta=6,8,10$:

\noindent\underline{$\Delta=6$}:
\bea
\mathcal{J}_\mu\mathcal{J}^\mu\,,&& c_{6,0}=-\frac{\Delta_L  (\Delta_L ^2-4 \Delta_L +9) (3 \tilde{f}_0-2 \tilde{h}_0)}{1680 (\Delta_L -3) (\Delta_L -2)}\cr
\mathcal{J}_\mu\mathcal{J}_\nu\,,&& c_{6,2}=-\fft{\tilde{f}_0\Delta_L(1+\Delta_L)}{560(\Delta_L-2)}\,, \label{d4c6-2}
\eea
\underline{$\Delta=8$}:
\bea
T_{\mu\nu}T^{\mu\nu}\,,&& c_{8,0}=\frac{\Delta_L }{201600 (\Delta_L -4) (\Delta_L -3) (\Delta_L -2)}
\Big(2 (\Delta_L  (\Delta_L  (\Delta_L  (7 \Delta_L -45)+100)
\cr && +100)+228) f_0^2-2 (\Delta_L  (\Delta_L  (\Delta_L  (7 \Delta_L -55)+130)+80)+168) f_0 h_0\cr
&&+40 \Delta_L  ((\Delta_L -3) \Delta_L +20) (2 f_4-h_4)+960 (2 f_4-h_4)\cr
&&+(\Delta_L -6) (\Delta_L  (\Delta_L  (7 \Delta_L -23)+22)+12) h_0^2\Big)\,,\cr
T_{\mu\rho}T_{\nu}^\rho\,,&& c_{8,2}=\frac{\Delta_L}{201600 (\Delta_L -3) (\Delta_L -2)}
\Big((21 \Delta_L ^3-49 \Delta_L ^2+126 \Delta_L +76) f_0^2\cr
&&-2 (7 \Delta_L ^3 -13 \Delta_L ^2+52 \Delta_L +32) f_0 h_0+80 (\Delta_L ^2+3 \Delta_L +2) f_4\Big)\,,\cr
T_{\mu\nu}T_{\rho\sigma}\,,&& c_{8,4}=\frac{\Delta_L (7 \Delta_L^2+6 \Delta_L +4) f_0^2}{201600 (\Delta_L -2)}\,,
\eea
\underline{$\Delta=10$}:
\bea
\mathcal{J}^\mu\mathcal{J}^\nu T_{\mu\nu}\,,&& c_{10,0}=\frac{\Delta_L}{2217600 (\Delta_L -5) (\Delta_L -4) (\Delta_L -3) (\Delta_L -2)}  \times \cr && \Big(2 f_0 ((\Delta_L  (\Delta_L  (\Delta_L  (\Delta_L  (11 \Delta_L -83)+281)+21)+1570)+1044) \tilde{h}_0
\cr && -(\Delta_L  (\Delta_L  (\Delta_L  (\Delta_L  (33 \Delta_L -199)+562)+1252)+6374)+4596) \tilde{f}_0)\cr
&&+11 \Delta_L ^5 h_0 (3 \tilde{f}_0-2 \tilde{h}_0) +\Delta_L ^4 (-199 \tilde{f}_0 h_0+186 h_0 \tilde{h}_0+80 h_6)\cr
&&+\Delta_L ^3 (793 \tilde{f}_0 h_0-582 h_0 \tilde{h}_0-80 h_6)+7 \Delta_L ^2 (259 \tilde{f}_0 h_0+94 h_0 \tilde{h}_0+400 h_6)\cr
&&+20 \Delta_L  (553 \tilde{f}_0 h_0-30 h_0 \tilde{h}_0+508 h_6)+144 (53 \tilde{f}_0 h_0-2 h_0 \tilde{h}_0+50 h_6)
\cr && -200 (\Delta_L +1) (\Delta_L +2) ((\Delta_L -4) \Delta_L +45) f_6\Big)\,,\cr
\mathcal{J}_\mu\mathcal{J}^\rho T_{\rho\nu}\,,&& c_{10,2}=\frac{\Delta_L}{2217600 (\Delta_L -4) (\Delta_L -3) (\Delta_L -2)} \times\cr
&& \Big(f_0 (2 (\Delta_L  (\Delta_L  (\Delta_L  (11 \Delta_L -6)+239)+374)+228) \tilde{h}_0\cr
&&-(\Delta_L  (\Delta_L  (\Delta_L  (99 \Delta_L -104)+927)+2024)+1224) \tilde{f}_0)\cr
&&+(\Delta_L +1) ((\Delta_L  (\Delta_L  (33 \Delta_L -1)+556)+720) \tilde{f}_0 \tilde{h}_0
\cr &&-200 (\Delta_L +2) (\Delta_L +3) f_6)\Big)\,,\cr
\mathcal{J}_\mu \mathcal{J}_\nu  T_{\rho\sigma}\,,&& c_{10,4}=-\frac{\Delta_L  (\Delta_L +1) (\Delta_L  (11 \Delta_L +21)+20) f_0 \tilde{f}_0}{739200 (\Delta_L -3) (\Delta_L -2)}\,.
\eea
While the detail can be complicated, the structures of non-vanishing OPE coefficients and the relevant exchange operators can be derived from the dimension analysis.  Up to and including $\Delta=10$, each $c_{\Delta, J}$ corresponds to one unique operator, the product of either $T_{\mu\nu}$ or ${\cal J}_\mu$. For $\Delta\ge 12$, $c_{\Delta, J}$ can have contributions from multiple operators, via the product of both $T_{\mu\nu}$ and ${\cal J}_\mu$. For example, to $c_{12,0}$, both ${\rm tr}(T^4)$ and ${\cal J}^2 {\cal J}^2$ can contribute. As was in the previous cases, the OPE coefficients here also involve integer poles of $\Delta_L$.  We shall comment this in section \ref{integerDelta}. Note that the lowest twisted OPE coefficients such as $c_{4n, 2n}$ that exist in the previous section remains the same, depending only on $f_0$, which have a universal CFT interpretation (\ref{f0}).  The new lowest-twisted OPE coefficients such as $c_{6,2}$ and $c_{10,4}$ depend also only and simply on $\tilde f_0$, analogous to the dependence of $c_{4n,2n}$ on $f_0$; however, $\tilde f_0$, being proportional to $Q^2$, does not have a clear CFT interpretation. This is one of the rather common features in the AdS/CFT correspondence where a simple bulk quantity does not lands itself as a straightforward parameter in the dual CFT.

\subsubsection{$d=6$}
In $d=6$, we shall present the lowest-twist results up to and including $\Delta=16$. This requires that metric functions $(h,f)$ expand to the order of $1/r^{16}$:
\be
f=1-\fft{f_0}{r^6}+\fft{\tilde{f}_0}{r^{10}}-\fft{f_6}{r^{12}}+\fft{f_8}{r^{16}}+\cdots\,,\qquad h=1-\fft{h_0}{r^6}+\fft{\tilde{h}_0}{r^{10}}-\fft{h_6}{r^{12}}+\fft{h_8}{r^{16}}+\cdots\,.
\ee
The OPE coefficients for the lowest-twisted operators of the type $c_{6n,2n}$ is the same as those in section 2 and they are universally depending on $f_0$, unaffected by the Maxwell fields.  However, including the Maxwell field in the construction of the bulk black hole does introduce new types of lowest-twisted operators. Here we present two explicit examples:
\bea
\mathcal{J}_\mu\mathcal{J}_\nu\,,&& c_{10,2}=-\fft{\tilde{f}_0\Delta_L(\Delta_L+1)(\Delta_L+2)}{11088(\Delta_L-4)(\Delta_L-3)}\,,\\
 \mathcal{J}_\mu \mathcal{J}_\nu  T_{\rho\sigma}\,,&& c_{16,4}=-\frac{\Delta_L  (\Delta_L +1) (\Delta_L +2) (221 \Delta_L ^3+1113 \Delta_L ^2+3286 \Delta_L +3360) f_0 \tilde{f}_0}{1372250880 (\Delta_L -6) (\Delta_L -5) (\Delta_L -4) (\Delta_L -3)}\,.\nn \label{d6c10-2}
\eea
These lowest-twist OPE coefficients depend only on $\tilde{f}_0$ and $f_0$.  In the next subsection, we show in general even $d$ dimensions that the lowest-twist OPE coefficients with at most two current operators $\mathcal{J}_\mu$ depend only on the mass parameter $f_0$ and the charge parameter $\tilde{f}_0$.

\subsection{Lowest-twist analysis}

We follow the analogous discussion in section 2. To isolate the lowest-twist contributions, we again take the large $\hat{u}$ limit while keeping $\xi=\hat{u}/r^{d/2}$ finite and non-vanishing. As seen in section 2, we hope this allows us to select all the lowest-twist contributions for the pure multi-stress tensor parts $T^n$ in arbitrary even $d$. Here, we consider conformal blocks with $n$ $T_{\mu\nu}$'s and $m$ $\mathcal{J}_\mu\mathcal{J}_\nu$'s, the lowest-twist contribution in $G(r,w,\hat{u})$ has the large-$r$ dependence
\be
G\sim\fft{\hat{u}^{2(m+n)}}{r^{dn+(d-1)m}}\sim \mathcal{O}(\fft{1}{r^{m(d-2)}})\,.
\ee
The contributions with higher twists for $\tilde{n}$ $T$ and $\tilde{m}$ $\mathcal{J}\mathcal{J}$ fall as
\be
G\sim\fft{\hat{u}^{2(\tilde{m}+\tilde{n}-k)}}{r^{d\tilde{n}+(d-1)\tilde{m}}}
\sim\mathcal{O}(\fft{1}{r^{\tilde{m}(d-2)+kd}})\,,\qquad k\geq1\,.
\ee
If we keep all orders up to $\mathcal{O}(\fft{1}{r^{m(d-2)}})$, then the condition that only the lowest twist contributions are preserved is
\be
m<\tilde{m}+\fft{d}{d-2}k\,.\label{con}
\ee
However, this condition cannot always be held.  For example, for $m=2$, we can easily find situations that violates the condition (\ref{con}), e.g. $\tilde{m}=0, k=1, d=4$. Thus we are not likely to isolate the lowest twist contributions. However, for $m=0$ and $m=1$, we find that (\ref{con}) is always satisfied. In other words, we can take large $\hat{u}$ limit and keep up to $\mathcal{O}(\fft{1}{r^{m(d-2)}})$ to isolate the lowest twist contributions with at most two current operators $\mathcal{J}_\mu\mathcal{J}_\nu$ involved.
Thus we shall consider two corresponding types of lowest-twist coefficients
\be
c_{\Delta=nd, J=2n}\,,\qquad \hbox{and}\qquad c_{\Delta=(n-1)d+2(d-1), J=2n}\,.\nn
\ee
It should be emphasized that this restriction arises only because we would like to give the result for general $d$. There is no such restriction if we consider a specific $d$.

In order to obtain the lowest-twist OPE coefficients with at most two current operators, we make the ansatz for $G$ as
\be
G(r,w,\hat{u})=Q^{(1)}(w,\xi)+\fft{Q^{(2)}(w,\xi)}{r^{d-2}}+\mathcal{O}(\fft{1}{r^n})\,,\qquad n>d-2\,.\label{ansatz-max-d}
\ee
Here $Q^{(1)}$ corresponds to $c_{nd,2n}$ and $Q^{(2)}$ corresponds to $c_{(n-1)d+2(d-1),2n}$. Substituting (\ref{ansatz-max-d}) into equation (\ref{eqG}), the leading order in the large-$r$ expansion gives equation for $Q^{(1)}(w,\xi)$ (\ref{reduced-eq-nomatter}) which was solved in \cite{Fitzpatrick:2019zqz}, (see section \ref{conc-G}.) The sub-leading order gives rise to the equation for $Q^{(2)}(w,\xi)$:
\bea
&&4\tilde f_0 \xi^2 \Big(4\Delta_L(\Delta_L+1) Q^{(1)} - w (4\Delta_L+1) \partial_w Q^{(1)} + w^2 \partial_w^2 Q^{(1)}\Big)\nn\\
&&+8 \Big((d-2)(\Delta_L-1) w^4 - 2(d-2) \Delta_L w^2 -2 \Delta_L(\Delta_L+1) \xi^2 f_0\Big) Q^{(2)} \nn\\
&&+\xi w^2 \big((d-2)(4\Delta_L+3d-10) w^2 - 8 d\Delta_L\big) \partial_\xi Q^{(2)} +
(d-2)^2 \xi^2 w^4 \partial_\xi^2 Q^{(2)}\nn\\
&&+4w \Big(f_0 \xi^2 (4\Delta_L +1) - w^2 \big( d(w^2-2) + (w^2-1)(2\Delta_L-5)\big)\Big) \partial_w Q^{(2)}\nn\\
&& + 4 \xi w^3 (d+2w^2 - dw^2) \partial_w\partial_\xi Q^{(2)} -4 w^2 (f_0 \xi^2 + w^2 - w^4)
\partial_w^2 Q^{(2)}=0\,.\label{eq-Q21}
\eea
Thus the general solution depends both $f_0$ and $\tilde f_0$.

\subsubsection{$n=1$}

The simplest case is $n=1$, corresponding to $J=2$.  There are two lowest-twist OPE coefficients $c_{d,2}$ and $c_{2(d-1),2}$. We make the ansatz
\be
Q^{(1)}=1+Q^{(1)}_2 \xi^2+\mathcal{O}(\xi^3)\,,\qquad Q^{(2)}=Q^{(2)}_2 \xi^2+\mathcal{O}(\xi^3)\,.
\ee
The solution for $Q^{(1)}_2$, corresponding to $c_{d,2}$ was obtained in \cite{Fitzpatrick:2019zqz}, and presented in section 2, depends only on $f_0$.  The quantity $Q^{(2)}_2$, corresponding to $c_{2(d-1),2}$, satisfies the equation
\bea
&& 2 w^3 ((d (4-3 w^2)+2 \Delta_L +(9-2 \Delta_L ) w^2-5) Q^{(2)'}_2(w)+w (w^2-1) Q^{(2)''}_2(w))
\cr && +4 Q^{(2)}_2(w) ((d-2) (d+2 \Delta_L -4)w^4-4 (d-1) \Delta_L  w^2)+8 \Delta_L  (\Delta_L +1) \tilde{f}_0=0\,.\label{eq-Q22}
\eea
Thus $Q^{(2)}_2$ depends only on $\tilde{f}_0$. To solve the equation (\ref{eq-Q22}), we adopt the polynomial ansatz
\be
Q^{(2)}(w)=\sum_{m=-m_d}^{m_u} q_m w^{m}\,,\qquad m\in 2\mathbb{Z}\,.
\ee
Subsequently, we have
\bea
&& 4\tilde{f}_0\Delta_L(\Delta_L+1)-\sum_{m=-m_d+2}^{m_u+2}(m-4d+2)(m-2(\Delta_L+1))q_{m-2}w^m
\cr && +\sum_{m=-m_d+4}^{m_u+4}(m-2d)(m-d-2\Delta_L)q_{m-4}w^m=0\,.
\eea
It follows from the analogous analysis below (\ref{ansatz3}), we see that the lower cutoff must be $-2$ and that the upper cutoff must be $2(d-2)$, i.e.
\be
m_d=-2\,,\qquad m_u=2(d-2)\,.
\ee
We now have a recursion relation for $q_m$
\be
q_{-2}=\frac{\Delta_L  \tilde{f}_0}{2 d-1}\,,\qquad
q_m = -\frac{q_{m-2} (2 d-m-2) (d+2 \Delta_L -m-2)}{(4 d-m-4) (m-2 \Delta_L )}\,,\qquad m\ge -1\,.
\ee
This can be solved straightforwardly, giving
\be
q_m=\fft{(d-2)\tilde{f}_0(2\Delta_L+d-4)(2\Delta_L+d-2)(-d+3)_{m/2-1}(-\Delta_L-\fft{d}{2}+3)_{m/2-1}}{8(2d-3)(2d-1)(\Delta_L-1)(-2d+4)_{m/2-1}
(-\Delta_L+2)_{m/2-1}}\,.
\ee
We find that the OPE coefficient $c_{2(d-1), 2}$ is
\bea
c_{2(d-1),2}=\ft{1}{4}q_{2(d-2)}
=\frac{ 2^{-2 d-1} \sqrt{\pi }\,\Gamma (d) \Gamma(1-\Delta _L) \Gamma(\frac{d}{2}-\Delta _L)}{\Gamma (d+\frac{1}{2}) \Gamma(-\frac{d}{2}-\Delta _L+1) \Gamma (d-\Delta _L-1)}\tilde{f}_0\,.
\eea
The $d=4$ ($c_{6,2}$) and $d=6$ ($c_{10,2}$) cases were already given in (\ref{d4c6-2}) and (\ref{d6c10-2}) respectively.

\subsubsection{Recursion relation for general $n$}

To obtain the OPE coefficients for higher conformal dimensions with at most two current operators, we take the ansatz (only for even $d$)
\be
Q^{(1)}=\sum_{m=-n}^{\fft{n(d-2)}{2}}\sum_{n=0}^{\infty} a_{nm}\xi^{2n}w^{2m}\,,\qquad
Q^{(2)}=\sum_{m=-n}^{\fft{(n+1)(d-2)}{2}}\sum_{n=0}^{\infty}b_{nm}\xi^{2n}w^{2m}\,.\label{Q1-Q2-all}
\ee
The upper bound in the polynomial truncation for $w$ is the conformal dimension. The recursion relation for all $a_{nm}$ was obtained in the previous section, see (\ref{recur-nomatter}). Substituting (\ref{Q1-Q2-all}) into (\ref{eq-Q21}) yields the following recursion relation for $b_{nm}$ with known $a_{nm}$
\bea
b_{nm}&=&\frac{1}{4 (\Delta_L-m ) (d n+d-m-2)}\Big(4 (\Delta_L-m ) (\Delta_L -m-1) (\tilde{f}_0\, a_{n-1,m+1}
\cr &&-f_0 q(n-1,m+1))\cr
&&+(d (n+1)-2 (m+n)) (2 \Delta_L+ (d-2) n -2 m)\,b_{n,m-1}\Big)\,,\label{recur-Q2}
\eea
together with
\bea
b_{0m}=0\,,\qquad b_{n,m<-n}=0\,,\qquad a_{n,m>\fft{n(d-2)}{2}}=0\,.\label{boundary-Q2}
\eea
Substituting (\ref{recur-nomatter}) into this recursion, we can evaluate $b_{n,m}$. We present two explicit low-lying examples:
\bea
&& d=4\,,\qquad b_{2,3}=-\frac{\Delta_L  (11 \Delta_L ^3+32 \Delta_L ^2+41 \Delta_L +20) f_0 \tilde{f_0}}{46200 (\Delta_L ^2-5 \Delta_L +6)}\,,\nn\\
&& d=6\,,\qquad b_{2,6}=-\frac{\Delta_L  (\Delta_L +1) (\Delta_L +2) (221 \Delta_L ^3+1113 \Delta_L ^2+3286 \Delta_L +3360) f_0 \tilde{f}_0}{85765680 (\Delta_L -6) (\Delta_L -5) (\Delta_L -4) (\Delta_L -3)}\,.\nn
\eea
The lowest OPE coefficients can be determined by $a_{nm}$ and $b_{nm}$ via
\bea
c_{\Delta=nd,J=2n} &=& \fft1{2^{J}}\,a_{n,\fft{d-2}{2}n}\quad\propto\quad f_0^n\,,\nn\\
c_{\Delta=(n-1)d+2(d-1), J=2n}&=&\fft1{2^J}\,{b_{n,\fft{(n+1)(d-2)}{2}}} \quad \propto\quad f_0^{n-1} \tilde f_0\,.
\eea

To conclude this section, we would like to remark that in order to obtain a general formula for all even $d$ dimensions, we restrict ourselves here to consider only the cases involving at most two current operators.  We also worked out a few examples in some specific low-lying dimensions.  Our results demonstrates that the OPE coefficients of the lowest-twisted operators involving $n_1$ stress tensor and $2n_2$ current operators is proportional to $f_0^{n_1} \tilde f_0^{n_2}$ with purely numerical coefficients, namely
\be
c_{\Delta, J} \propto f_0^{n_1} \tilde f_0^{n_2}\,, \qquad \hbox{for}\qquad
\Delta=n_1 d + 2 n_2 (d-1)\,,\quad J=2n_1 + 2n_2\,.
\ee

\section{The case of integer $\Delta_L$}
\label{integerDelta}

In the previous sections, we saw that the OPE coefficients can be divergent when $\Delta_L$ is an integer.
For a generic minimally-coupled free scalar discussed in the previous sections, we can avoid dealing with this problem by considering only the non-integer $\Delta_L$.  However, in gauged supergravity theories, scalars are typically conformally massless, corresponding to integer $\Delta_L=d-2$.  (The conformal dimensions for massive scalars are typically integers as well.) Since there is a great motivation to compute the holographic OPE coefficients from gauged supergravities, in this section we study the integer $\Delta_L$ case.

\subsection{Anomalous conformal dimensions}

For integer $\Delta_L$, the short set $G^s$ alone diverges, we therefore have to take the double-trace set into account to fix the divergence. In general, the double-trace set, as $\Delta_L$ approaching a certain integer $n$, the $(w/r)^{2\Delta_L}$ factor of $G^L(r,w,\hat u)$ in (\ref{ansatz1}) will have an extra logarithmic term, namely
\be
\big(\fft{w}{r}\big)^{2\Delta_L}\qquad\rightarrow\qquad \big(\fft{w}{r}\big)^{2n}\Big(1+2(\Delta_L-n)\log\big(\fft{w}{r})\Big)\,,
\ee
such that the factor $(\Delta_L-n)$ in front of $\log\big(\fft{w}{r}\big)$ will cancel the pole $1/(\Delta_L-n)$ in $G^L_i$ and leave us a finite logarithmic term in the near-boundary expansion for $G$. We thus take the ansatz
\bea
&& G(r,w,\hat{u})=\sum_i \fft1{r^i}\Big(G_i^1(r,w,\hat{u})+G_i^2(r,w,\hat{u})\log r\Big)\,,
 \cr && G_i^j=\sum_n^{i-m}\sum_m(\alpha^{ij}_{nm}+\beta^{ij}_{nm}\log w)w^n\hat{u}^m\,,\qquad j=1,2\,.\label{ansatz-log}
\eea
with $i$ taking all admissible power values and $(\alpha,\beta)$'s are constants. Thus for integer $\Delta_L$, we can no longer decompose $G$ into $G^T$ and $G^L$, but instead into terms with log and terms without log.
Note that here we shall not explicitly present the lower and upper bound of the polynomial expansions, with the understanding that a consistent solution will have an automatic truncation. We have verified with many examples that the ansatz (\ref{ansatz-log}), after plugging into equation (\ref{eqG}), can indeed be solved without any poles. However, not all the constants $(\alpha,\beta)$'s can be determined. In particular, those that contribute to the OPE coefficients involving the double-trace set are now typically undetermined. In fact these are the integration constants that require imposing further boundary conditions to fix \cite{Fitzpatrick:2019zqz}. As the result, even if we simply ignore all the logarithmic terms, the OPE coefficients, involving the double trace operators, can not be totally determined.

On the other hand, we find that the all the coefficients of $\log(z\bar z)$ can be fully determined by the equations of motion. To understand the the physical meaning of these logarithmic terms, it is worth noting that when $\Delta_L$ is an integer, taking $r\rightarrow\infty$, the $\log r$ terms are not suppressed, and their coefficients are actually the conformal anomalies. For the convergent finite terms in the large-$r$ expansion, in addition to polynomials of $z$ and $\bar{z}$ that appear in conformal blocks, there are now terms with overall $\log (z\bar{z})$ and they do not appear in conformal block. We take the view that these $\log(z\bar{z})$ terms should be interpreted as anomalous dimensions for the exchanged operators of bare conformal dimension $\Delta$.\footnote{We thank Kuo-Wei Huang for suggesting this interpretation.} Recall the conformal blocks $G^{00}_{\Delta,J}$ in the OPE limit \cite{Fitzpatrick:2019zqz}, for small anomalous dimension $\epsilon$, we have
\be
G^{00}_{\Delta+\epsilon,J}=G^{00}_{\Delta,J}\Big(1+\ft12\epsilon \log(z\bar{z})+\mathcal{O}((\epsilon\log z\bar z)^2)\Big)\,.
\ee
Thus it is advantageous to define
\be
\delta G_{\Delta, J}^{00} = \lim_{\epsilon\rightarrow 0} \fft{dG_{\Delta+\epsilon,J}^{00}}{d\epsilon}\,.
\ee
The holographic dictionary (\ref{holodict}) (dropping the $\log r$ conformal anomalous terms) now becomes
\bea
\lim_{r\rightarrow \infty} G(r,w,\hat u) &=&\sum_{\mathcal{O}}\left(c_{\Delta,J}\,G^{0,0}_{\Delta,J}(z,\bar{z}) +
\gamma_{\Delta, J}\,  \delta G^{0,0}_{\Delta,J}(z,\bar{z})\right)\cr
&=& \sum_{\mathcal{O}}c_{\Delta,J}\left(1+ \epsilon(\Delta, J) \delta G_{\Delta, J}^{0,0}\right)\,,\qquad
\epsilon(\Delta,J) \equiv \fft{\gamma_{\Delta, J}}{c_{\Delta,J}}\,.
\label{holodict2}
\eea
As we shall see from an explicit example in the next subsection, the coefficients $\gamma_{\Delta, J}$ can be completely determined from the asymptotic structure whilst the coefficients $c_{\Delta, J}$ are not without imposing the boundary condition in the middle of the spacetime.

\subsection{An explicit demonstration}

As a concrete example, we consider $\Delta_L=2$ in $d=4$, corresponding to a conformally massless scalar. It also saturates the BF bound.  Since the procedure of this construction was not covered in \cite{Fitzpatrick:2019zqz}, we shall give a detail demonstration. We consider the metric ansatz (\ref{expansion-max-4d}) up to and including $1/r^6$. The relevant expansion of $G(r,w,\hat{u})$ at large $r$ is
\be \label{max-spec1}
G(r,w,\hat{u})=1+\frac{G^1_4(w,\hat{u})+G^2_4(w,\hat{u})\log(r)}{r^4}
+\frac{G^1_6(w,\hat{u})+G^2_6(w,\hat{u})\log r}{r^6} + \cdots\,,
\ee
where
\bea
&&
G^1_4(w,\hat{u})=\sum^{4-m}_{n=-2}\sum^{2}_{m=0}(\alpha_{nm}^{41}+\beta_{n m}^{41} \log w)w^n \hat{u}^m  \, ,
\cr &&
G^2_4(w,\hat{u})=\sum^{4-m}_{n=-2}\sum^{2}_{m=0}(\alpha_{nm}^{42}+\beta_{n m}^{42} \log w)w^n \hat{u}^m  \, ,
\cr &&
G^1_6(w,\hat{u})=\sum^{6-m}_{n=-4}\sum^{4}_{m=0}(\alpha_{nm}^{61}+\beta_{n m}^{61} \log w)w^n \hat{u}^m  \, ,
\cr &&
G^2_6(w,\hat{u})=\sum^{6-m}_{n=-4}\sum^{4}_{m=0}(\alpha_{nm}^{62}+\beta_{n m}^{62} \log w)w^n \hat{u}^m  \, ,
\eea
Substituting these into the equation of motion (\ref{eqG}) for $G(r,w,\hat u)$, we find that the equation can be solved at the $1/r^6$ order, and the solutions for $(\alpha, \beta)$ can be obtained.  The non-vanishing coefficients are
\bea
&&\alpha_{-20}^{41}=-\frac{2}{5}(f_0+h_0)\, , \qquad \alpha_{00}^{41}=\frac{1}{20}(7f_0+h_0)\, , \qquad \alpha_{20}^{41}=\frac{1}{30}(f_0+2h_0)\, ,\nn\\
&&\alpha_{-22}^{41}=-\frac{2f_0}{5}\,,\qquad \alpha_{02}^{41}=-\frac{f_0}{5}\, , \qquad \alpha_{22}^{41}=-\frac{f_0}{15}\,,\nn\\
&&
\alpha_{-20}^{61}=\frac{2(\tilde{f}_0+\tilde{h}_0)}{7}\, ,\qquad \alpha_{00}^{61}=\frac{-7\tilde{f}_0-\tilde{h}_0}{28}\, , \qquad
\alpha_{20}^{61}=\frac{-12 \tilde{f}_0+5\tilde{h}_0}{140}\, ,\nn\\
&&\alpha_{-22}^{61}=\frac{2\tilde{f}_0}{7} \, , \qquad \alpha_{02}^{61}=\frac{3\tilde{f}_0}{14}\,, \qquad
\alpha_{22}^{61}=\frac{6\tilde{f}_0}{35}\, , \qquad
\beta_{40}^{61}=\frac{3\tilde{f}_0-5\tilde{h}_0}{70}\,,\nn\\
&&\beta_{60}^{61}=\frac{6\tilde{f}_0+5\tilde{h}_0}{210}\, ,\qquad
\beta_{42}^{61}=-\frac{3\tilde{f}_0}{35}\, ,\nn\\
&&
\alpha_{60}^{62}=\frac{1}{210}(-3\tilde{f}_0+20\tilde{h}_0-
210\alpha_{40}^{62}+210\alpha_{40}^{62}+630\alpha_{60}^{61}+315\alpha_{42}^{61})\,,\nn\\
&&
\alpha_{42}^{62}=\frac{1}{150}(3\tilde{f}_0-20\tilde{h}_0+140\alpha_{40}^{62}-
210\alpha_{40}^{61}-630\alpha_{60}^{61}-315\alpha_{42}^{61})\,.
\eea
We find that the coefficients $(\alpha^{40}_{41}, \beta^{40}_{42})$ and
$(\alpha^{40}_{61}, \alpha^{60}_{61}, \alpha^{42}_{61}, \alpha^{40}_{62})$ are arbitrary and can not be determined by the equation.  All the other coefficients vanish.  Substituting the solutions back to $G(r,w, \hat u)$ where $(w,\hat u)$ are defined by (\ref{change-va}), we find that the large-$r$ expansion has logarithmic divergence proportional to $\log r$.  As we have discussed earlier, these are related to the conformal anomaly.  The finite part is
\bea
&&\lim_{r \rightarrow \infty}G(r,w,\hat{u})=\cr
&&\qquad\frac{f_0}{60}z \bar{z}(z^2-2 z \bar{z}+\bar{z}^2)+z^2 \bar{z}^2\alpha^{40}_{41}+z^3\bar{z}^3\alpha^{60}_{61}-\frac{1}{4}z^2\bar{z}^2(z^2-2z \bar{z}+\bar{z}^2)\alpha^{42}_{61}\nn\\
&&\qquad
+\Big(\frac{1}{30}(f_0-h_0)+\frac{1}{840}(9 \tilde{f}_0 z^2-6\tilde{f}_0 z \bar{z}+10 \tilde{h}_0 z \bar{z}+9 \tilde{f}_0 \bar{z}^2)\Big) z^2\bar z^2\log(z\bar{z}).
\eea
To apply the holographic dictionary (\ref{holodict2}), we note that
\bea
&& G_{4,0}^{0,0} = z^2\bar z^2\,,\quad G_{4,2}^{0,0}=z\bar{z}(z^2+z \bar{z}+\bar{z} z +\bar{z}^2)\,,\quad
G_{4,4}^{0,0}=z^4+z^3\bar{z}+z^2\bar{z}^2+z \bar{z}^3+\bar{z}^4\,,\cr
&&G_{6,0}^{0,0}=z^3 \bar{z}^3\,,\quad G_{6,2}^{0,0}=z^2\bar{z}^2(z^2+z \bar{z}+\bar{z}^2)\,,\quad G_{6,4}^{0,0}=z\bar{z}(z^4+z^3\bar{z}+z^2\bar{z}^2+z\bar{z}^3+\bar{z}^4)\,,\cr
&&G_{6,6}^{0,0}=z^6+z^5\bar{z}+z^4\bar{z}^2+z^3\bar{z}^3+z^2\bar{z}^4+z^2\bar{z}^4+z\bar{z}^5+\bar{z}^6\,,
\eea
and that
\be
\delta G_{\Delta, J}^{0,0} = \ft12 G_{\Delta,J}^{0,0} \log(z\bar z)\,.
\ee
We can thus read off all $c_{\Delta, J}$ and $\gamma_{\Delta, J}$ coefficients from the first line in
(\ref{holodict2}), and the non-vanishing ones are
\bea
&&\gamma_{4,0} = \fft{f_0-h_0}{15}\,,\qquad \gamma_{6,0}=\frac{1}{84}(-3\tilde{f}_0+2 \tilde{h}_0)\,, \qquad \gamma_{6,2}=\frac{3\tilde{f}_0}{140}\,,\\
&& c_{4,0}=\frac{1}{20}(-f_0+20 a^{40}_{41})\,, \quad c_{4,2}=\frac{f_0}{60}\,,\quad
c_{6,0}=\frac{1}{4}(4\alpha^{60}_{61}+3\alpha^{42}_{61})\, , \quad c_{6,2}=-\frac{\alpha^{42}_{61}}{4}.\nn
\eea
Thus we see that all the $\gamma_{\Delta, J}$'s are determined by the asymptotic structures of the black hole solution, but not all the $c_{\Delta, J}$'s, which are expected. We present some more low-lying examples of $\gamma_{\Delta, J}$. For $d=4$ and $\Delta_L=2$, the non-vanishing coefficients are
\bea
\gamma_{6,0}&=&\fft{1}{84}(-3\tilde{f}_0+2\tilde{h}_0)\,,\qquad \gamma_{6,2}=\fft{3\tilde{f}_0}{140}\,,\nn\\
\gamma_{8,0} &=&\fft{-29f_0^2+26f_0 h_0+2(-60f_4+h_0^2+30h_4)}{2520}\,,\nn\\
\gamma_{8,2}&=&\fft{15f_0^2+48f_4-14f_0 h_0}{2520}\,,\qquad  \gamma_{8,4}=-\fft{11f_0^2}{12600}\,,\nn\\
\gamma_{10,0}&=&\fft{-2472 f_0\tilde{f}_0-2952 f_6+2061 \tilde{f}_0 h_0+554 f_0\tilde{h}_0-62h_0\tilde{h}_0}{166320}\,,\nn\\
\gamma_{10,2} &=&-\fft{2433 f_0\tilde{f}_0+3000 f_6-1569 \tilde{f}_0 h_0-1030 f_0\tilde{h}_0}{277200}\,,\quad
\gamma_{10,4}=-\fft{53f_0\tilde{f}_0}{30800}\,.
\eea
For $d=6$ and $\Delta_L=4$, we present some lowest-twist results
\be
 \gamma_{10,2}=\fft{5\tilde{f}_0}{231}\,,\quad \gamma_{12,4}=-\fft{-397f_0^2}{300300}\,,
\quad \gamma_{16,4}=\fft{2019f_0\tilde{f}_0}{476476}\,.
\ee
It is clear that the $\gamma_{\Delta, J}$ coefficients for the lowest-twist operators also depend only on the $f_0$ and $\tilde f_0$.

\subsection{An alternative derivation}

It is actually straightforward to understand why $\gamma_{\Delta,J}$'s could all be determined, since they can be derived directly from the $c_{\Delta, J}$'s discussed earlier.  Recall that for the short set alone, the OPE coefficients have poles for integers $\Delta_L=n$ behaving like
\be
c^s_{\Delta,J}=\fft{A(\Delta_L)\Delta_L}{\cdots(\Delta_L-n-1)(\Delta_L-n)(\Delta_L-n+1)\cdots}
\,,\label{pole-like}
\ee
where $A(\Delta_L)$ is a regular function of $\Delta_L$ without poles. When $\Delta_L$ is an integer, $c^{L}_{\Delta,J}$ for the light exchange operators must mix with $c^s_{\Delta,J}$ with some specific $(\Delta,J)$. As we have seen in the previous subsection,  we may adopt (\ref{ansatz-log}) to obtain OPE coefficients $c_{\Delta,J}$ which can be decomposed into $c^s_{\Delta,J}+c^L_{\Delta,J}$. The full OPE coefficients $c_{\Delta,J}$ should be smooth for integer $\Delta_L=n$. We can simply write $c^L_{\Delta,J}$ in terms the regular function $c_{\Delta,J}$ as
\be
c^L_{\Delta,J}=-c^s_{\Delta,J} U(\Delta_L)+c_{\Delta,J}\,,\label{cphi}
\ee
where $U|_{\Delta_L\rightarrow n} \rightarrow 1$.  It follows from (\ref{ansatz1}) that for $\Delta_L=n$ we have a prefactor in front of double-trace modes
\be
w^{2\Delta_L}=w^{2n}(1+2(\Delta-n)\log w)\sim (z\bar{z})^{2n}(1+2(\Delta-n)\log (z\bar{z}))\,,
\ee
hence for double-trace operators we must have following terms
\be
c^L_{\Delta,J}(1+2(\Delta-n)\log(z\bar{z}))\,.
\ee
Now we can conclude the coefficients $\gamma_{\Delta,J}$'s can also be readily read off as
\be
\gamma_{\Delta,J}=\lim_{\Delta\rightarrow n}(2(\Delta-n)c^L_{\Delta,J})\,.
\ee
Applying (\ref{cphi}), $c_{\Delta,J}$ is a regular function and so it is removed by the limit $\lim_{\Delta\rightarrow n}$, we thus end up with
\be
\gamma_{\Delta,J}=-2\lim_{\Delta\rightarrow n}((\Delta-n)c^s_{\Delta,J})=-2{\rm Res}_{\Delta_L=n}c^s_{\Delta,J}\,.\label{rela}
\ee
For those $c^s_{\Delta,J}$ without the pole $\Delta_L=n$, even though some $c^L_{\Delta,J}$ will still mix with it, there will be no poles involved, then the factor $(\Delta_L-n)$ in $\gamma_{\Delta,J}$ simply suppresses it gives rise to $\gamma_{\Delta,J}=0$. Therefore the formula (\ref{rela}) is still valid.

In the previous section we adopted the dictionary (\ref{holodict2}) to derive the low-lying $\gamma_{\Delta, J}$  for $d=4$ and $\Delta_L=2$ and $d=6$, $\Delta_L=6$.  It is easy to verify that these results can also be simply obtained from the explicit $c_{\Delta, J}$ of general $\Delta_L$ in section \ref{OPEcharge}.  Thus many properties of $c_{\Delta,J}$'s will be inherited by the corresponding $\gamma_{\Delta, J}$. In particular,
the $\gamma_{\Delta,J}$ coefficient of the lowest-twisted operators involving $n_1$ stress tensor and $2n_2$ current operators is proportional to $f_0^{n_1} \tilde f_0^{n_2}$ with purely numerical coefficients, namely
\be
\gamma_{\Delta, J} \propto f_0^{n_1} \tilde f_0^{n_2}\,, \qquad \hbox{for}\qquad
\Delta=n_1 d + 2 n_2 (d-1)\,,\quad J=2n_1 + 2n_2\,.
\ee

\section{Gauged supergravity examples}
\label{gauge-sg}

From the analysis in section \ref{comments}, we find $c_{d,0}\propto f_0-h_0$. For pure gravity black holes where $f_0=h_0$, it follows that $c_{d,0}=0$, which is consistent with $T_\mu^\mu=0$. On the other hand, black holes with $f_0\neq h_0$ do exist and hence $c_{d,0}\neq0$. In this case, there are other operators with conformal dimension $\Delta=d$ and spin $J=0$. It is then of great interest to consider such $f_0\ne h_0$ black holes that are supported by appropriated matter fields. In the previous section, we consider charged black holes that the condition $f_0=h_0$ continues to hold. Although new conserved current operator with conformal dimension $\Delta=d-1$ is introduced, it can by no means contribute $c_{d,0}$, leading to a consistent result.

In this section, we consider AdS black holes in gauged supergravities where additional matter fields are involved. In particular, we consider charged AdS black holes in STU models which can be embedded in M-theory or type IIB strings as rotating branes via Kaluza-Klein sphere reductions \cite{Cvetic:1999xp}.

The STU model was originally \cite{Duff:1995sm} referred to the ${\cal N}=2$ consistent truncation of ${\cal N}=8$, $D=4$ supergravity in such a way that the S-duality, T-duality and U-duality all emerge. In the gauged version, the $SO(8)$ gauged group of gauged maximal supergravity is reduced to the $U(1)^4$ subgroup. The terminology of the STU model was later generalized to include all the maximum number of $U(1)$ subgroup truncations of maximal gauged supergravities in higher dimensions. Since our construction requires the boundary dimensions to be even, we consider gauged supergravities in $D=5$, 7 dimensions and then consider supergravity inspired models in all even dimensions.

\subsection{$D=5$ STU models}

\subsubsection{${\rm U}(1)^3$ gauged supergravity}

${\cal N}=8$, $D=5$ gauged supergravity allows to have a consistent and supersymmetric $U(1)^3$ truncation, which sometimes is referred to as the $D=5$ STU model. We follow the notation of \cite{Cvetic:1999xp} and write the Lagrangian for the bosonic sector as ${\cal L}=\sqrt{-g} L$, where
\bea
&& L=R-V(\varphi_1,\varphi_2)-\ft{1}{2}(\partial\varphi_1)^2-\ft{1}{2}(\partial\varphi_2)^2-\ft{1}{4}\sum_{i=1}^3 X_i^{-2}(F^i)^2+\ft{1}{4}\epsilon
^{\mu\nu\rho\sigma\lambda}F^1_{\mu\nu}F^2_{\rho\sigma}A^3_\lambda\,,
\cr && V(\varphi_1,\varphi_2)=-4\ell^{-2}\sum_{i=1}^3 X_i^{-1}\,,\qquad X_i=e^{-\fft{1}{2}\vec{a}_i\cdot\vec{\varphi}}\,,\qquad \vec{\varphi}=(\varphi_1,\varphi_2)\,,
\cr && \vec{a}_1=(\fft{2}{\sqrt{6}},\sqrt{2})\,,\qquad \vec{a}_2=(\fft{2}{\sqrt{6}},-\sqrt{2})\,,\qquad \vec{a}_3=(-\fft{4}{\sqrt{6}},0)\,.\label{STU5}
\eea
The Lagrangian admits charged AdS planar black hole \cite{Behrndt:1998jd,Cvetic:1999xp}. In the Euclidean signature, the solution is given by
\bea
&& ds_5^2=(H_1H_2H_3)^{-\fft{2}{3}}\tilde f dt^2+(H_1H_2H_3)^{\fft{1}{3}}(\tilde f^{-1}d\rho^2+\rho^2d\Omega_{3,0}^2)\,,
\cr && X_i=H_i^{-1}(H_1H_2H_3)^{\fft{1}{3}}\,,\qquad A^i=\sqrt{\ft{\mu}{q_i}} (1-H_i^{-1})dt\,,
\cr && \tilde f=-\fft{\mu}{\rho^2}+\rho^2 H_1H_2H_3\,,\qquad H_i=1+\fft{q_i^2}{\rho^2}\,,\label{STU5-bh-1}
\eea
where we set the AdS radius $\ell$ to unity. To proceed on computing the holographic OPE coefficients, we first express the solution (\ref{STU5-bh-1}) in the form of (\ref{metric}).  There is no close such a form, we present the metric in the large-$r$ expansion.  Define
\be
r^2=(H_1H_2H_3)^{\fft{1}{3}}\rho^2\,,
\ee
for large $r$, we have
\bea
h &=& -\fft{g_{tt}}{r^2} =1 - \fft{\mu}{r^4} + \fft{\mu q^\1}{3 r^6} -
\fft{\mu q^\2}{9 r^8} + \fft{\mu q^\3}{81 r^{10}} + \cdots\,,\nn\\
f &=& \fft{r^2}{g_{\rho\rho}} \big(\fft{dr}{d\rho}\big)^2 =1 -\fft{\mu -
\fft29 q^\2}{r^4} + \fft{\ft13 \mu q^\1 -\fft4{81} q^\3}{r^6} -
\fft{q^\2(\fft13\mu -\fft1{27} q^\2)}{r^8} \nn\\
&&+\fft{\fft1{81}\mu (6 q^\1 q^\2 + 5 q^\3) - \fft4{729} q^\2 q^\3}{r^{10}} + \cdots\,,
\eea
where we denote
\bea
q^\1 &=& q_1 + q_2 + q_3\,,\qquad q^\2=q_1^2 + q_2^2 + q_3^2 - q_1 q_2 - q_1 q_3 - q_2 q_3\,,\cr
q^\3&=&(2q_1 - q_2 - q_3)(2q_2 - q_1 - q_3) (2q_3 - q_1 - q_2)\,.
\eea
Thus the expansions of $f$ and $h$ have exactly the same form as the general Maxwell case (\ref{expansion-max-4d}), specializing in $d=4$. However, we now must have $f_0\ne h_0$, since the scalar fields  $(\varphi_1,\varphi_2)$ are involved in the black hole solution.  The scalar fields will be turned off if we set all $q_i$ equal, in which case, the solution reduces to the RN-AdS black hole, with $h=f$.

Having obtained the large-$r$ behavior of the charged black hole, we can follow the previous sections and consider a free scalar $\phi$ and obtain the two-point function associated with this free scalar in the black hole background. It is clear that the formulae obtained in sections \ref{OPE} (for $d=4$) apply, and no further new OPE coefficients could emerge. Even though in this case, the result appears to be exactly the same as in section \ref{OPEcharge} for $d=4$, we now have $c_{4,0}\ne 0$ owing to the scalar contribution to the metric such that $h_0\ne f_0$. Thus there must be $\Delta=4$ and $J=0$ operators in exchange.  These are precisely supplied by the scalar operators dual to $(\varphi_1,\varphi_2)$.  To see this, we note that asymptotically, both $(\varphi_1,\varphi_2)$ vanish and their scalar potential, up to and including the quadratic order, is
\be
V=-12-2(\varphi_1^2+\varphi_2^2)+\cdots\,,
\ee
In other words, the scalars have the mass and dual conformal dimensions
\be
m_1^2=m_2^2=-4\,,\qquad \Delta_1=\Delta_2=2\,.
\ee
Therefore, the spectrum contains two different scalar operators with the same conformal dimension $\Delta=2$, saturating the BF bound. They can contribute to the OPE coefficient $c_{4,0}$ in the form of $\mathcal{O}_1 \mathcal{O}_1$, $\mathcal{O}_2 \mathcal{O}_2$ and $\mathcal{O}_1 \mathcal{O}_2$. When $q_1=q_2=q_3$, for which both $(\varphi_1,\varphi_2)$ vanish, and $h_0=f_0$, in which case, $c_{4,0}=0$ and consistent with the fact that now there is no $\Delta=4$, $J=0$ operator in exchange, since the only possible candidate $T^\mu_\mu$ vanishes for any CFT in the flat background.

   Although everything discussed above appears to be consistent, the picture remains somewhat unsatisfactory.
The first is that the free scalar is not part of the STU model, but introduced by hand.  The second is related to the observation that there is no single scalar OPE coefficient, i.e. $c_{2,0}$ which could be contributed by either $\mathcal{O}_1$ or $\mathcal{O}_2$. The vanishing of $c_{2,0}$ here is not itself inconsistent with the conformal blocks, but highly coincidental. In fact, both issues can be resolved within the STU model itself.

\subsubsection{${\rm U}(1)^2$ truncation and perturbation}

In the previous subsection, we introduced a free scalar propagating on the charged black hole in the STU model. The free scalar however lies outside of the STU model.  In this subsection, we consider the scalar perturbation within the STU model, to examine whether the non-vanishing $c_{2,0}$ can emerge.  The general perturbation is very complicated and we consider a special case.  We truncate the $U(1)^3$ system to a $U(1)^2$ system by setting two out of three Maxwell fields equal, namely $A^1=A^2\equiv A/\sqrt2$, in which case $\varphi_2=0$ decouples from the charged black hole.  The metric of the resulting black hole can be simply obtained by setting $q_2=q_1$ in (\ref{STU5-bh-1}), which means we have $H_1=H_2$. The large-$r$ expansion has the same form, but the specific coefficients in each falloffs are specialized to $q_2=q_1$.

Instead of introducing a free scalar outside the theory, we start with the above reduced background with $\varphi_2=0$ and consider the linear perturbation
\be
\varphi_2=\bar{\varphi}_2+\phi\,,\qquad \bar{\varphi}_2=0\,.\label{pert-STU}
\ee
We find that the linearized equation is
\be
\Big(\Box+4e^{\fft{1}{\sqrt{6}}\varphi_1}-\ft{1}{2}F^2 e^{\fft{2}{\sqrt{6}}\varphi_1}\Big)\phi=0\,,\label{phi2}
\ee
where the quantities in the bracket are the solutions of the reduced $U(1)^2$ theory. Compared to the free scalar equation (\ref{phi1}) where the mass is a constant, the ``mass'' in (\ref{phi2}) is now $r$-dependent, and we may write (\ref{phi2}) as
\be
(\Box-m(r)^2)\phi=0\,.\label{phi3}
\ee
The leading term of the large-$r$ expansion of $m(r)^2$ is the constant mass squared which gives rise to the conformal dimension $\Delta_L$. The first few terms for the expansion of $m(r)^2$ are
\be
m^2(r)=\Delta_L(\Delta_L-4)-\fft{m_0}{r^2}-\fft{m_2}{r^4}-\fft{m_4}{r^6}-\fft{m_6}{r^8}-\cdots\,,
\label{D5stumrsq}
\ee
where
\bea
\Delta_L&=&2\,,\quad m_0 = \ft43 (q_1-q_3)\,,\quad
m_2 = - \ft49 (q_1-q_3)^2\,,\cr
m_4 &=& \ft8{81} (q_1^3-3 q_1^2 q_3+3 q_1 ( q_3^2+27\mu)-q_3^3)\,,\quad
m_6=\ft{16}{3} \mu q_1 (q_1-q_3)\,.
\eea
The leading term tells us that $\Delta_L=2$ and it is an integer. Thus in this case, as was discussed in section \ref{integerDelta}, most of OPE coefficients are undetermined using the holographic procedure. We can however derive the anomalous dimension related coefficients $\gamma_{\Delta,J}$, following the exact procedure outlined in section \ref{integerDelta}. We find some low-lying examples:
\bea
\gamma_{4,0}&=&\frac{16(f_0-h_0)-5(3m_0^2+4 m_2)}{240}\,,\quad \gamma_{6,2}=\frac{18\tilde{f}_0-7 f_0 m_0 }{840}\,,\quad  \gamma_{8,4}=-\frac{11 f_0^2}{12600}\,,\cr
\gamma_{6,0} &=& \frac1{6720}\big(-240\tilde{f}_0+160\tilde{h}_0+7(-24f_0 m_0 +16 h_0 m_0+5m_0^3-20 m_0 m_2+32 m_4)\big)\, ,\cr
\gamma_{8,0} &=&\frac{1}{645120}\Big(-7420f_0^2+512h_0^2+384(-80f_4+40h_4+3\tilde{f}_0m_0-5\tilde{h}_0m_0)\cr
&&+32h_0(-21m_0^2+4m_2) +64 f_0(104h_0+21 m_0 + 21m_0^2+52m_2)\cr
&&-7(15m_0^4+120m_0^2 m_2+112 m_2^2+384m_0 m_4)\Big)\,,\cr
\gamma_{8,2}&=&\frac{120f_0^2+384f_2-112f_0h_0-108\tilde{f}_0 m_0+21 f_0 m_0 +52 f_0 m_2}{20160}\,.
\label{D5stugamma}
\eea

It should be noted that the large-$r$ expansion of $m(r)^2$ contains the $1/r^2$ term, implying that
the ansatz for $G$ must also include the quadratic power $1/r^2$, which gives rise to non-vanishing $c_{2,0}$. To obtain $c_{2,0}$ and also verify (\ref{D5stugamma}) using the general formula (\ref{rela}), we turn to consider (\ref{D5stumrsq}) with a general $\Delta_L$. The large-$r$ expansion for $f$ and $h$ is assumed to take the same form as in (\ref{expansion-max-4d}). Following the same procedure, we find
\bea
c_{2,0} &=& \fft{m_0}{4(\Delta_L-1)}\,,\qquad c_{4,2}=\fft{f_0\Delta_L}{120}\,,\qquad
c_{6,2}=\frac{\Delta _L (7 f_0 m_0-6 \tilde{f}_0 (\Delta _L+1))}{3360 (\Delta _L-2)}\,,
\cr &&\cr
c_{4,0} &=&\frac{4 (\Delta_L -1) ((\Delta_L -4) \Delta_L  (f_0-h_0)+5 m_2)+15 m_0^2}{480 (\Delta_L -2) (\Delta_L -1)}\,,
 \cr &&
 \cr c_{6,0} &=&\frac{1}{13440 (\Delta _L-3) (\Delta _L-2) (\Delta _L-1)}\Big(28 m_0 (\Delta _L-1) (f_0 (\Delta _L-5) \Delta _L
\cr && -h_0 (\Delta _L-4) \Delta _L+5 m_2)-8 (\Delta _L-1) \Delta _L (3 \tilde{f}_0 ((\Delta _L-4) \Delta _L+9)\cr
&&-2 (\tilde{h}_0 ((\Delta _L-4) \Delta _L +9) +7 m_4))+35 m_0^3\Big)\,,
 \cr &&
 \cr c_{8,0} &=& \frac{1}{3225600 (\Delta _L-4) (\Delta _L-3) (\Delta _L-2) (\Delta _L-1)}\Big(\cr
 && 840 m_0^2 (\Delta _L-1) (\Delta _L (f_0 (\Delta _L-6)-h_0 (\Delta _L-4))+5 m_2)\cr
 &&+16 [10 m_2 (\Delta _L-1) \Delta _L (f_0 (\Delta _L-6) (7 \Delta _L-1)+h_0 ((25-7 \Delta _L) \Delta _L
-24))\cr
&&+\Delta _L (-2 f_0 h_0 (\Delta _L-1) (\Delta _L (\Delta _L (\Delta _L (7 \Delta _L-55)+130)+80)+168)
 \cr && +(\Delta _L-1) (80 f_4 (\Delta _L+1) ((\Delta _L-4) \Delta _L+24)\cr
&&+h_0^2 (\Delta _L-6) (\Delta _L (\Delta _L (7 \Delta _L-23)+22)+12)
 \cr && -40 (\Delta _L+1) (h_4 ((\Delta _L-4) \Delta _L+24)-9 m_6))\cr
 &&+2 f_0^2 ((\Delta _L (7 \Delta _L-52)+145) \Delta _L^3+128 \Delta _L-228))
 +35 m_2^2 (\Delta _L-1) (5 \Delta _L-3)]\cr
 &&-480 m_0 (\Delta _L-1) \Delta _L (\Delta _L ((3 \tilde{f}_0-2 \tilde{h}_0) \Delta _L-15 \tilde{f}_0+8 \tilde{h}_0)
 \cr && +6 (4 \tilde{f}_0-3 \tilde{h}_0)-14 m_4)+525 m_0^4\Big)\,,
 \cr c_{8,2}&=&\frac{\Delta _L}{403200 (\Delta _L-3) (\Delta _L-2)} \Big(20 (\Delta _L+1) (8 f_4 (\Delta _L+2)-9 \tilde{f}_0 m_0)\cr
 &&+f_0 (5 (4 m_2 (7 \Delta _L-1) +21 m_0^2)-4 h_0 (\Delta _L (\Delta _L (7 \Delta _L-13)+52)+32))\cr
 &&+2 f_0^2 (7 \Delta _L (\Delta _L (3 \Delta _L-7)+18)+76)\Big)\,,
 \cr c_{8,4}&=&\frac{f_0^2 \Delta _L (\Delta _L (7 \Delta _L+6)+4)}{201600 (\Delta _L-2)}\,.\label{STU-gene}
\eea
Note that the OPE coefficients $c_{2,0}$ and $c_{4,2}$ remains convergent when $\Delta_L=2$.  The other OPE coefficients become undetermined, as was discussed in section \ref{integerDelta}. It can be easily verified that the $\gamma$ coefficients (\ref{D5stugamma}) can indeed be obtained from (\ref{STU-gene}) using the formula (\ref{rela}).
The non-vanishing of $c_{2,0}$ gives a more consistent picture that there is an exchange operator of $\Delta=2$ in the spectrum. However, the procedure has the shortcoming in dealing with gauged supergravity models since the scalar fields are typically conformally massless with integer $\Delta_L$.

\subsection{$D=7$ STU model}

\subsubsection{${\rm U}(1)^2$ gauged supergravity}

Seven-dimensional gauged supergravity from the Kaluza-Klein $S^4$ reduction of $D=11$ supergravity have a consistent $U(1)^2$ truncation, the relevant bosonic Lagrangian is
\bea
&& L=R-V(\varphi_1,\varphi_2)-\ft{1}{2}(\partial\varphi_1)^2-\ft{1}{2}(\partial\varphi_2)^2-\fft{1}{4}\sum_{i=1}^2 X_i^{-2}(F^i)^2\,,\qquad
X_i=e^{-\fft{1}{2}\vec{a}_i\cdot\vec{\varphi}}\,,
\cr && V(\varphi_1,\varphi_2)=-16 X_1 X_2-8 X_1^{-1}X_2^{-2}-8 X_2^{-1}X_1^{-2}+4(X_1 X_2)^{-4}\,,\quad \vec{\varphi}=(\varphi_1,\varphi_2)\,,
\cr && \vec{a}_1=(\sqrt{2},\sqrt{\ft{2}{5}})\,,\qquad \vec{a}_2=(-\sqrt{2},\sqrt{\ft{2}{5}})\,,\label{STU7}
\eea
The charged AdS planar black hole was given in \cite{Cvetic:1999xp}.  In Euclidean signature, it is
\bea
&& ds_7^2=(H_1H_2)^{-\fft{4}{5}}\tilde f dt^2+(H_1H_2)^{\fft{1}{5}}(\tilde f^{-1}d\rho^2+\rho^2dx^i dx^i)\,,
\cr && X_i=H_i^{-1}(H_1H_2)^{\fft{2}{5}}\,,\qquad A^i=\sqrt{\mu/q_i}(1-H_i^{-1})dt\,,
\cr && \tilde f=-\fft{\mu}{\rho^4}+\rho^2 H_1H_2\,,\qquad H_i=1-\fft{q_i}{\rho^4}\,,\label{STU7-bh-1}
\eea
The large-$r$ expansion for $f$ and $h$ up to $1/r^{10}$ in the metric coordinate choice (\ref{metric}) is
\bea
&& f=1-\fft{\mu}{r^6}-\fft{2\mu(q_1+q_2)}{5r^{10}}+\cdots\,,
\cr && h=1-\fft{\mu}{r^6}+\fft{2(3q_1^2-4q_1 q_2+3q_2^2)}{25r^8}-\fft{2\mu(q_1+q_2)}{5r^{10}}+\cdots\,.
\eea
Note in this case we have $f_0=h_0$ and $\tilde{f}_0=\tilde{h}_0$. There is no new power lower than $1/r^6$; however, a new term of $1/r^8$ now appears in $h$. This term is contributed by the two scalars and reflects the non-vanishing results of $c_{8,0}$. Indeed, from the potential in (\ref{STU7}), it is easy to see that
\be
V=-30-4(\varphi_1^2+\varphi_2^2)+\cdots\,,
\ee
reflecting there are two operators with conformal dimensions $\Delta=4$. The coefficient $c_{8,0}$ will be contributed by the products $\mathcal{O}_1^2$, $\mathcal{O}_2^2$ and $\mathcal{O}_1\mathcal{O}_2$. Thus the spectrum contains light conformal primary operators of the energy-momentum tensor, conserved current and scalars, with $(\Delta, J)=(6,2), (5,1)$ and $(4,0)$, respectively. There is no composite of these operators that could give $(6,0)$. It follows that the additional operators in $D=7$ STU model have no $c_{6,0}$ contribution, due to the highly coincidental $f_0=h_0$ in the black holes solutions of $D=7$ STU model. (The situation becomes clearer in section \ref{scalarhair}.) Nevertheless, we still present the more interesting case with $m(r)^2$ as in the $D=5$ STU model, and we will present the OPE coefficients up to $c_{10,J}$.

\subsubsection{The $U(1)$ truncation}

The $D=7$ STU model (\ref{STU7}) can be further truncated to a smaller sector with only a $U(1)$ symmetry. This consistent truncation can be done by setting $\varphi_1=0$ and $F^1=F^2=F/\sqrt{2}$. Then we consider the black hole solution of this $U(1)$ theory as the background where $q_2=q_1$.  The linearized equation for $\varphi_1$ then becomes (\ref{phi3}) with
\be
m(r)^2=-8e^{\fft{3}{\sqrt{10}}\varphi_2}+\fft{1}{2}F^2 e^{\sqrt{\fft{2}{5}}\varphi_2}\,,
\ee
where $\varphi_2$ takes the solution of the truncated theory. The first few expansions are
\be
m(r)^2=-8+\frac{24 q_1}{5 r^4}-\frac{72 q_1^2}{25 r^8}+\cdots\,.\label{m-STU7-ex}
\ee
We now have $\Delta_L=4$, thus there is no more contribution to $c_{d,0}$. On the other hand, it is still of value to present results and verify (\ref{rela}). We take the general ansatz for $D=7$ as follows
\bea
&& m^2(r)=\Delta_L(\Delta_L-6)-\fft{m_0}{r^4}-\fft{m_2}{r^6}-\fft{m_4}{r^8}-\fft{m_6}{r^{10}}-\cdots\,,
\cr && f=1-\fft{f_0}{r^6}+\fft{\tilde{f}_0}{r^{10}}-\fft{f_6}{r^{12}}\cdots\,,\qquad h=1-\fft{h_0}{r^6}+\fft{h_\phi}{r^8}+\fft{\tilde{h}_0}{r^{10}}-\fft{h_6}{r^{12}}\cdots\,.
\eea
Note $m_2=0$ in the $D=7$ $U(1)$-truncated STU model, see (\ref{m-STU7-ex}). Nevertheless, we keep it there for generality and present the results:
\bea
c_{4,0} &=&\fft{m_0}{24(\Delta_L-1)}\,,\quad c_{6,0}=\frac{\left(f_0-h_0\right) \left(\Delta _L-6\right) \Delta _L+7 m_2}{840 \left(\Delta _L-3\right)}\,,
\cr &&
\cr c_{6,2} &=&\fft{f_0\Delta_L}{560}\,,\quad c_{8,0}=\frac{8 (\Delta _L-1) \Delta _L (h_{\phi } ((\Delta _L-6) \Delta _L+20)+9 m_4)+7 m_0^2 (5 \Delta _L-8)}{40320 (\Delta _L-4) (\Delta _L-3) (\Delta _L-1)}\,,
\cr &&
\cr   c_{10,0}&=&\frac{1}{3326400 (\Delta _L-5) (\Delta _L-4) (\Delta _L-3)}\Big(55 m_0 (f_0 \Delta _L (3 \Delta _L^2-26 \Delta _L+16)
\cr && -3 h_0 \Delta _L (\Delta _L^2-6 \Delta _L+8)+3 m_2 (7 \Delta _L-8))-40 \Delta _L (\Delta _L+1) (5 \tilde{f_0} (\Delta _L^2-6 \Delta _L+50)
\cr && -3 \tilde{h_0} (\Delta _L^2-6 \Delta _L+50)-33 m_6)\Big) \cr &&
\cr  c_{10,2}&=&\frac{\Delta _L (55 f_0 m_0 (3 \Delta _L-2)-200 \tilde{f}_0 (\Delta _L^2+3 \Delta _L+2))}{2217600 (\Delta _L-4) (\Delta _L-3)}\,.
\eea
When $\Delta_L=4$, as in the supergravity case, $c_{4,0}$, $c_{6,0}$ and $c_{6,2}$ remains finite, but others become divergent and undetermined. Using the procedure outlined in section \ref{integerDelta}, we find that all the $\gamma_{\Delta, J}$ coefficients however are fully determined, with the non-vanishing ones given by.
\bea
\gamma_{8,0} &=& -\frac{-96h_\phi-7 m_0^2 -72 m_4}{5040}\,,\qquad \gamma_{10,2}=\frac{120\tilde{f}_0-11 f_0 m_0}{5544}\,,\nn\\
\gamma_{10,0}&=&\frac{-1680\tilde{f}_0+1008 \tilde{h}_0+11(8f_0 m_0-3m_0 m_2-24 m_6)}{16632}\,,\nn\\
\gamma_{12,0}&=&\frac{1}{51891840}(9 (16 f_0 (9902 h_0+1183 m_2)-199536 f_0^2
\cr && +12272 h_0 m_2+19184 h_0^2-2431 m_2^2)-10 (708480 f_6
\cr && +13 (1008 m_0 h_{\phi }+11 m_0^3+396 m_4 m_0+4320 m_8)-354240 h_6))\,,\nn\\
\gamma_{12,2} &=& \frac{f_0 \left(5915 m_2-38240 h_0\right)+47832 f_0^2+105000 f_6}{3003000}\,,\qquad \gamma_{12,4}=\frac{-397 f_0^2}{300300}\,.
\eea
These coefficients can also be obtained by using the relation (\ref{rela}).

\subsection{Supergravity inspired models}

Inspired by supergravity, Chow proposed \cite{Chow:2011fh}  a class of Einstein-Maxwell-Dilaton models in general dimensions that generalized $D=5$ STU model (\ref{STU5}) and $D=7$ STU model (\ref{STU7}). The Lagrangian is
\bea
L&=&R-V-\fft{1}{2}(\partial\varphi_1)^2-\fft{1}{2}(\partial\varphi_2)^2-\fft{1}{4}\sum_{i=1}^2 X_i^{-2}(F^i)^2\,,\qquad
X_i=e^{-\fft{1}{2}\vec{a}_i\cdot\vec{\varphi}}\,,
\cr V&=&-(D-3)^2 X_1 X_2-2(D-3)(X_1 X_2)^{-\fft{D-3}{2}}(X_1+X_2)\cr
&&+(D-5)(X_1 X_2)^{-(D-3)}\,,
\cr
\vec{\varphi}&=&(\varphi_1,\varphi_2)\,,\qquad \vec{a}_1=(\sqrt{\ft{2}{D-2}},\sqrt{2})\,,\qquad \vec{a}_2=(\sqrt{\ft{2}{D-2}},-\sqrt{2})\,.\label{chow}
\eea
When $D=7$, it recovers $D=7$ gauged STU model (\ref{STU7}).\footnote{The convention is somewhat different from (\ref{STU7}) where $\varphi_1$ and $\varphi_2$ interchanges.} When $D=5$, (\ref{chow}) can give us $D=5$ supergravity (\ref{STU5}) with $A^3=0$. Charged AdS planar black holes can be obtained by simply generalizing $D=5$ (\ref{STU5-bh-1}) and $D=7$ (\ref{STU7-bh-1}), namely
\bea
ds_D^2 &=&(H_1H_2)^{-\fft{D-3}{D-2}}\tilde f dt^2+(H_1H_2)^{\fft{1}{D-2}}(\tilde f^{-1}d\rho^2+\rho^2 dx^i dx^i)\,,\nn\\
X_i &=& H_i^{-1}(H_1H_2)^{\fft{D-3}{2(D-2)}}\,,\qquad A^i=\sqrt{\fft{\mu}{q_i}} (1-H_i^{-1})dt\,,\nn\\
 \tilde f &=& -\fft{\mu}{\rho^{D-3}}+\rho^2 H_1H_2\,,\qquad H_i=1-\fft{q_i^2}{\rho^{D-3}}\,,\qquad i=1,2\,.\label{chow-bh}
\eea
In this subsection, we only focus on the case with $r$ dependent mass $m(r)^2$. As in the previous STU models,  we can consistently truncate $\varphi_2=0$ by requiring $F^1=F^2=F/\sqrt{2}$. The resulting theory admits the the black holes in (\ref{chow-bh}) with $q_1=q_2$.  We then turn on the linear perturbation (\ref{pert-STU}) and obtain the linearized scalar equation (\ref{phi3}) with
\be
m^2(r)=-2(D-3)e^{\fft{D-4}{\sqrt{2(D-2)}}\varphi_1}+\fft{1}{2}e^{\sqrt{\fft{2}{D-2}}\varphi_1}F^2\,.
\label{chow-mass}
\ee
For this generalized model in general dimensions $d$, we are only concerned about what is the new OPE coefficient below $\Delta=d$. We thus substitute (\ref{chow-bh}) with $q_1=q_2=q$ and (\ref{chow-mass}) into (\ref{phi3}), we find the first two terms are as follows
\bea
m^2(r)=-2(d-2)-\fft{m_0}{r^{d-2}}+\cdots\,,\qquad m_0=-\fft{2(d-2)(d-3)q}{d-1}\,.\label{m-chow}
\eea
Thus we conclude that we have $\Delta_L=d-2$, and the lowest conformal block has conformal dimensions $\Delta=d-2$.

We now turn to compute $c_{d-2,0}$ with a general $\Delta_L$. Since now the spin is zero, we are allowed to take the simplest ansatz for $G$ as
\be
G(r,w,\hat{u})=1+\fft{G^{(d-2)}(w)}{r^{d-2}}+\cdots\,.\label{ansatz-chow}
\ee
Substituting (\ref{ansatz-chow}) and the expansion (\ref{m-chow}) into equation (\ref{eqG}), the reduced equation is
\bea
&& -2(d-2)(w^2(\Delta_L-1)-2\Delta_L)G^{(d-2)}+w((d(w^2-2)+(w^2-1)(2\Delta_L-5))G^{(d-2)'}
\cr && -2(m_0+(w^2-1)G^{(d-2)''}))
=0\,.\label{reduce-eq-chow}
\eea
The equation (\ref{reduce-eq-chow}) admits following exact solution
\bea
&& G^{(d-2)}=\fft{1}{2(\Delta_L-1)\Gamma(d-1)}\Big(m_0 w^{d-4}(w^2-1)^{1-\fft{d}{2}}(w^d \Gamma(\fft{d}{2}-1)\Gamma(\fft{d}{2})
\cr && -w^2 \Gamma(d-2))\,_2F_1[2-\ft12{d},\ft{1}{2}(d-2);\ft12{d};{w^{-2}}]\Big)\,.\label{G-d-2}
\eea
It follows that the exact solution (\ref{G-d-2}) is a finite polynomial of $w$ only for even $d$, while for odd $d$ it is a infinite series of $w$. Actually, all contributions $G^i$ of $G$ have this property, and the exact solutions for higher order contributions are hard to come by. For this reason, as mentioned in section \ref{OPE}, we only consider even $d$ such that we have polynomial ansatz. Given (\ref{G-d-2}), we have
\be
c_{d-2,0}=\fft{m_0 \Gamma(\fft{d}{2}-1)\Gamma(\fft{d}{2})}{2(\Delta_L-1)\Gamma(d-1)}=-\fft{2q \Gamma(\fft{d}{2})^2}{\Gamma(d)}\,.
\ee
The consistency condition requires that the spectrum should have a scalar operator with conformal dimension $\Delta$ in a such way this scalar can contribute $c_{d,0}$ whenever $f_0\neq h_0$. However, for the theory (\ref{chow}) we consider in this subsection, the lowest conformal dimension $\Delta=d-2$ can contribute to $c_{d,0}$ if and only if $d=4$.  From the black hole solutions (\ref{chow-bh}), it is clear that we can have $f_0\neq h_0$ if and only if $d=4$.

\section{Scalar hairy AdS black holes and the issue of $c_{d,0}$}
\label{scalarhair}

In section \ref{OPE}, we examined the holographic approach to calculate the the OPE coeficients in the conformal black blocks of the four-point scalar functions in a dual $d=$even dimensional CFT. The calculation was based on the assumption that the large-$r$ falloffs of the metric functions took the form (\ref{hfexpansion}). This assumption and the condition $h_0=f_0$ were ensured as long as the gravity sector involves only the graviton, with no other massive modes.

In this section,  we study $c_{d,0}$ in the context of general classes of AdS scalar hairy black holes.  Recently large classes of exact hair black holes were constructed, see, e.g.~\cite{Anabalon:2012ta,
Anabalon:2013qua,Anabalon:2013sra,Gonzalez:2013aca,Feng:2013tza,Fan:2015oca}. Exact time-dependent solutions describing the formation of black holes were also constructed
\cite{Zhang:2014sta,Lu:2014eta,Xu:2014xqa,Fan:2015tua,Fan:2015ykb}.  In this paper, however, an exact solution is not required in our analysis. It is clear that the OPE coefficient $c_{d,0}$ can be contributed by a single operator ${\cal O}$ of $\Delta=d$ or by the product ${\cal O}_1{\cal O}_2$ with $\Delta_{1,2}=d/2$.  In the latter case, we then should expect that a new coefficient $c_{d/2,0}$ emerges in general.  We shall discuss in detail how these issues arise and could be resolved in scalar hairy black holes.

\subsection{AdS scalar hairy black holes and their asymptotic structures}

For simplicity, we consider only one scalar $\Phi$ that is involved in the construction of the static AdS planar black holes.  The relevant part of the Lagrangian in $D=d+1$ dimensions is
\be
{\cal L}_{d+1}=\sqrt{-g} (R - \ft12 (\partial\Phi)^2 - V(\Phi) + \cdots)\,,\label{genscalarlag}
\ee
where the ellipses denote additional matter or curvature terms that can be involved in the solution but that do not affect our leading or sub-leading falloffs of the static black hole metric (\ref{metric}). Furthermore, the ellipses also include a new scalar $\phi$ whose linearized equation of motion in the black hole background takes the general form (\ref{phi3}).

We first study the property of the scalar $\Phi$.  Assuming that $V(\Phi)$ has a fixed point $\Phi=0$ and the theory admits the AdS vacuum of radius $\ell=1$.  For small $\Phi$, we expect that $V(\phi)$ has the Taylor expansion
\be
V(\Phi) = V(0) + \ft12 m^2 \Phi^2 + \gamma_3 \Phi^3 + \gamma_4 \Phi^4 + \cdots\,,
\ee
where $m^2=\Delta(\Delta-d)$ and $\gamma_3, \gamma_4$, etc.~are all constants. In gauged supergravities, $\gamma_3$ typically vanishes and $\Delta=d-2$ and the corresponding $\Phi$ is conformally massless.
The BF bound requires that $\Delta\ge \ft12 d$ and in $d=4$, the conformally massless scalar saturates the BF bound. Note that if an AdS planar black hole contains the scalar $\Phi$ hair, then we expect that there is an exchange operator ${\cal O}_\Phi$ of $\Delta$ and $J=0$ that contribute the relevant OPE coefficients.

We now consider AdS planar black holes involving $\Phi$.  The large-$r$ expansions of $(h,f)$ and $\Phi$ can then be determined \cite{Lu:2014maa,Liu:2015tqa}. The leading and sub-leading terms are
\bea
\Phi &=& \fft{\Phi_1}{r^{d-\Delta}} + \cdots + \fft{\Phi_2}{r^{\Delta}} + \cdots\,,\nn\\
f &=& 1 - \fft{f_0}{r^d} + \cdots + \fft{c_1\Phi_2^2}{r^{2\Delta}} + \cdots\,,\nn\\
h &=& 1 + \fft{d-\Delta}{2(d-1)} \fft{\Phi_1^2}{r^{2(d-\Delta)}} + \cdots - \fft{h_0}{r^d}+ \cdots +
\fft{c_2\Phi_2^2}{r^{2\Delta}} + \cdots\,,\nn\\
\eea
where $\Phi_1$ and $\Phi_2$ are constants. The $\Phi_2^2$ terms above are determined by dimensional analysis; therefore, the dimensionless coefficients $(c_1,c_2)$ remains to be determined by a specific theory. The $\Phi_1^2$ term in $h$ is fully determined, by the equation
\be
\Phi'^2 = - \fft{(d-1) }{r} \fft{h}{f}\big(\fft{f}{h}\big)'\,.\label{phieom}
\ee
This equation is generally true provided that $\Phi$ is the only scalar mode involved in the black hole.

At the first sight, the asymptotic structure appears to suggest that there are four independent integration constants $(h_0,f_0,\Phi_1, \Phi_2)$, the equations of motion reduce them to three.  Furthermore, the existence of an event horizon for being a black hole reduces further to two.  In most of exact solutions, either $\Phi_1$ or $\Phi_2$ vanishes, (see STU black holes in the previous section.) When $(\Phi_1,\Phi_2)$ are both non-vanishing, subtleties emerge in the first law of black hole dynamics \cite{Liu:2013gja,Lu:2014maa}. The only known exact solution of this type is the Kaluza-Klein dyonic AdS black hole \cite{Lu:2013ura} and its generalization \cite{Chow:2013gba}.

\subsection{$\Delta=d$}

A natural candidate as an exchange operator for non-vanishing $c_{d,0}$ is a scalar operator with $\Delta=d$.  If $\Phi$ is precisely the bulk dual of such an operator, then $\Phi$ is massless ($m^2=0$), but not conformally massless. The asymptotic expansion for $\Phi$ takes the form
\be
\Phi= \Phi_1 + \fft{\Phi_2}{r^d} + \cdots\,.
\ee
It's back reaction to the metric functions are of higher orders such that $h\sim 1-h_0/r^d + \cdots$ and $f\sim 1 -f_0/r^d + \cdots$.  Thus for minimally coupled scalar $\phi$ with (\ref{phi3}), where $m(r)$ is a constant, a non-vanishing $c_{d,0}$ would naturally arise without introducing any issues.

We may also consider the possibility that the light operator $\phi$ couples to $\Phi$, in the following way
\be
{\cal L}=\sqrt{-g} \Big(-(\partial\Phi)^2 - (\partial\phi)^2  - \ft12 u(\Phi) \phi^2 + \cdots\Big)\,.
\ee
This types of coupling is inspired by the STU models discussed earlier.  If the function $u(\Phi)$ for small $\Phi$ expands as
\be
u(\Phi) = m_0^2 + \alpha_1 \Phi + \alpha_2 \Phi^2 + \cdots\,,\label{uphi}
\ee
then at large-$r$ expansion, the $m(r)^2$ in (\ref{phi3}) behaves as
\be
m(r)^2 = m_0^2 + \fft{\alpha_1 \Phi_2}{r^d} + \cdots\,.
\ee
Now in addition to $f_0=h_0$, $c_{d,0}$ also depends on $\alpha_1 \Phi_1$. However, since $\Phi$ is not conformally massless, it does not typically arise in gauged supergravities.

\subsection{$\Delta=d/2$}

When $\Delta=d/2$, the mass of $\Phi$ saturates the BF bound and the scalar is conformally massless in only $d=4$. The large-$r$ expansion for $\Phi$ is
\be
\Phi=\fft{\Phi_1 \log r + \Phi_2}{r^{\fft12 d}} + \cdots\,.
\ee
We require that $\Phi_1=0$ for the AdS black hole.  In this case, we no longer have $f_0=h_0$, but instead we have $f_0-h_0\sim \Phi_2^2$. Now for $d\ge 6$, the holographic OPE coefficient $c_{d,0}$ depends not only on $f_0$, but also on the $\Phi_2^2$.

Introducing $\Phi$ with $\Delta=d/2$ to the black hole would imply the possibility of a new OPE coefficient $c_{d/2,0}$.  This coefficient will be zero holographically if we consider the minimally coupled scalar $\phi$.
In order to generate a non-vanishing $c_{d/2,0}$, it is necessary to consider the coupling between $\Phi$ and $\phi$, e.g.
\be
{\cal L}=\sqrt{-g} \Big(-(\partial\Phi)^2 - (\partial\phi)^2  +\ft14 d^2 \Phi^2 - \ft12 u(\Phi) \phi^2 + \cdots\Big)\,,
\ee
where $u(\Phi)$ takes the same form as (\ref{uphi}).  Then we have $m(r)^2\sim m_0^2 + \alpha_1 \Phi_2/r^{d/2} + \cdots$ in the large-$r$ expansion.
It would be instructive to present $c_{d/2,0}$ explicitly. Assuming the metric does not have power $1/r^{d/2}$, the equation can be solved for even $d/2$ and we end up with
\be
c_{\fft{d}{2},0}=\frac{\alpha _1 \Phi _2(-1)^{\frac{d}{4}+1} 2^{-\frac{d}{2}-1} \Delta   \Gamma \left(\frac{1}{2}-\frac{d}{4}\right) \Gamma \left(\frac{d}{4}\right) \Gamma (-\Delta ) \Gamma \left(\frac{d}{2}-\Delta \right)}{\sqrt{\pi } \Gamma \left(\frac{d}{4}-\Delta +1\right)^2}\,.\label{cdover2}
\ee
For instance, for $d=4$, (\ref{cdover2}) immediately gives rise to $c_{2,0}$ in (\ref{STU-gene}) for the $D=5$ STU model.

\section{Conclusions}
\label{conc}

In this paper, we studied the holographic OPE coefficients for heavy-light scalar four-point functions in the heavy limit where the heavy scalars approximately create a black hole background $\mathcal{O}_H|0 \rangle\simeq |{\rm BH}\rangle $. For black holes constructed by pure (massless) gravity sector, we constructed the OPE coefficients $c_{d,0}$ in general even $d$. The OPE coefficient $c_{d,0}$ is proportional to $f_0-h_0$ which is always zero in pure gravity black hole involving only the massless graviton. This is consistent with the fact that $T_\mu^\mu=0$ for CFTs in flat spacetime.  We then studied black holes involving matter fields that admit the possibility for $f_0\neq h_0$ and hence necessarily exhibit more operators in the spectrum of the dual CFTs.

We included the Maxwell field and considered charged AdS black holes in a general class of gravity-Maxwell theories. The Maxwell field can contribute the conserved current operator $\mathcal{J}$  with $\Delta=d-1,J=1$ to exchange in the conformal blocks in the boundary CFT. The explicit low-lying OPE coefficients in $d=4$ and $d=6$ were presented. The recursion formula for the lowest-twist OPE coefficients involving at most two current operators were obtained. Our investigation indicates that the lowest-twist OPE coefficients associated with the charged black hole takes the form $c_{\Delta=n_1 d + 2 n_2 (d-1), J=2n_1 + 2n_2} \propto f_0^{n_1} \tilde f_0^{n_2}$ where $f_0$ and $\tilde{f}_0$ are related to the black hole mass and charge respectively. However, the conserved current operator $\mathcal{J}$ is not lying in the track of $c_{d,0}$, which is consistent with the fact that charged black holes remain $f_0=h_0$.

Motivated by the fact that scalars in supergravities are typically conformally massless with $\Delta_L=d-2$, we studied the OPE coefficients when $\Delta_L$ is an integer. In this case, the solutions of the linearized scalar equation of the light operator involve logarithmic dependence and we presented a detail procedure to read off the coefficients $\gamma_{\Delta,J}$. Even though the OPE coefficients $c_{\Delta, J}$ can not be fully determined, the coefficients $\gamma_{\Delta,J}$ that are related to the anomalous dimensions can nevertheless be. In addition, we presented a general residue formula for extracting $\gamma_{\Delta,J}$ from $c_{\Delta, J}$ with generic $\Delta_L$. For the charged black holes in gravity-Maxwell theories discussed in section 3, we find that for the lowest-twist operators we have $\gamma_{\Delta=n_1 d + 2 n_2 (d-1), J=2n_1 + 2n_2} \propto f_0^{n_1} \tilde f_0^{n_2}$.

We then investigated the charged AdS black holes in $D=5,7$ gauged supergravity STU models and their generalization in general dimensions. These black holes not only involve multiple Maxwell fields, but also a set of scalar fields. As was mentioned earlier, the scalars are conformally massless and are dual to operators with $\Delta=d-2$. In addition to following the earlier example and introducing a free scalar as the light operator, we consider linear perturbation of one of the scalars in the STU supergravity models.  This allows to
discuss the holographic properties within the context of supergravities.  We obtained the OPE coefficient $c_{d-2,0}$ explicitly. In $D=5$, $d=4$, although $f_0\neq h_0$ and $c_{d,0}\neq0$ owing to the scalar contribution, the results are consistent, since in $d=4$, conformally massless scalars have $\Delta_L=2$, and hence a product of two of such scalar operators can contribute $c_{d,0}$. For $\Delta_L=d-2$, the coefficients $\gamma_{\Delta,J}$ were also presented and verify the formula (\ref{rela}).

We analyzed the generic scalar falloffs in asymptotic AdS geometry. We found that $c_{d,0}\neq0$ is not rare in the framework of scalar hairy black holes when the scalars that are dual to operators with $\Delta=d$ or $\Delta=d/2$ are involved in the black hole construction.

Our preliminary investigation of the gauged STU models in section \ref{gauge-sg} indicates that the general procedure of the holographic OPE coefficients of heavy-light four-point functions can be analysed within the framework of supergravities. The price however is that $\Delta_L$ is now an integer such that the OPE coefficients are not fully determined. It should be emphasized that massive scalar modes also arise in supergravities in the Kaluza-Klein spherical reductions, but again they generally have integer $\Delta_L$. For example, it follows from the general scalar formula in appendix A of \cite{Bremer:1998zp} that the (massive) breathing modes in sphere reductions of M-theory or type IIB string give rise to massive scalars with $\Delta_L=12, 10,8,6$ in $d=6,5,4,3$ respectively. Scalars with non-integer $\Delta_L$ are hard to come by if not entirely impossible in the consistent truncation of gauged supergravities involving massive modes. Thus the interior boundary conditions in all these cases must be required to constrain the linearized solution. However, the framework adopted in this paper is based on the near-boundary expansion. It is thus of great interest to develop new techniques to relate the interior data to the asymptotic values.

\section*{Acknolwedgement}

We are grateful to Kuo-Wei Huang for clarifying for us the key points in their paper \cite{Fitzpatrick:2019zqz} at the early stage of this work. We are grateful to Jun-Bao Wu for useful discussion. We are also grateful to the JHEP referee for pointing out a serious technique error regarding to $c_{d,0}$ in the earlier version of the paper. The work is supported in part by NSFC (National Natural Science Foundation of China) Grants No.~11875200 and No.~11475024.

\appendix

\section{Conformal blocks}
\label{conformalblock}

Conformal blocks capture the essence of the four-point functions in conformal field theories. In this appendix, we present some properties of conformal blocks of scalar four-point functions. By the virtual of the conformal symmetry, the four-point functions can be written in a compact form
\be
\langle\mathcal{O}_1(x_1)\mathcal{O}_2(x_2)\mathcal{O}_3(x_3)\mathcal{O}_4(x_4)\rangle=
\fft{g(u,v)}{(x_{12}^2)^{\fft{\Delta_1+\Delta_2}{2}}(x_{34}^2)^{\fft{\Delta_3+\Delta_4}{2}}}
\Big(\fft{x_{24}^2}{x_{14}^2}\Big)^{\fft{\Delta_{12}}{2}}
\Big(\fft{x_{14}^2}{x_{13}^2}\Big)^{\fft{\Delta{34}}{2}}\,,\label{4pt}
\ee
where $x_{ij}=x_i-x_j$, $\Delta_{ij}=\Delta_i - \Delta_j$ and $g(u,v)$ is a function of the cross ratios $(u,v)$:
\bea
 u=\fft{x_{12}^2 x_{34}^2}{x_{13}^2 x_{24}^2}=z\bar{z}\,,
 \qquad v=\fft{x_{14}^2 x_{23}^2}{x_{13}^2 x_{24}^2}=(1-z)(1-\bar{z})\,.
\eea
To study the four-point functions, it is standard and convenient to use the conformal symmetry to take the conformal frame, namely
\be
x_1=(0,0,\cdots)\,,\qquad x_2=(x,y,0,\cdots)\,,\qquad x_3=(1,0,\cdots)\,,\qquad x_4\rightarrow\infty \,.\label{cf}
\ee
Defining $z=x+{\rm i}y$ and $\bar z=x-{\rm i} y$, we have
\be
u=z\bar z\,,\qquad v=(1-z)(1-\bar z)\,.
\ee
It is in this conformal frame which we wrote (\ref{heavy-light-4pt}) in section \ref{OPE}.

By applying the OPE expansion, $g(u,v)$ is expected to be decomposed into conformal blocks characterized by conformal dimension $\Delta$ and spin $J$
\be
g(u,v)=\sum_{\Delta,J}\lambda^J_{12\Delta}\lambda^J_{34\Delta}
G^{\Delta_{12},\Delta_{34}}_{\Delta,J}(z,\bar{z})\,,
\ee
where $\lambda_{ij\Delta}$'s are the coefficients in OPE expansions and hence the three-point functions are $\langle \mathcal{O}_i\mathcal{O}_j\mathcal{O}_{\Delta,J}\rangle \propto\lambda^J_{ij\Delta}$. Throughout this paper, we actually denote $c_{\Delta,J}=\lambda^J_{12\Delta}\lambda^J_{34\Delta}$ and call it an OPE coefficient. It should be thus understood that $c_{\Delta, J}$ is not only a function of $(\Delta, J)$, but also $\Delta_{12}$ and $\Delta_{34}$.

The conformal block $G^{\Delta_{12},\Delta_{34}}_{\Delta,J}(z,\bar{z})$ is the eigenfunction of the quadratic Casimir invariant with respect to conformal algebra, namely
\bea
\mathcal{C}_2 \,G^{\Delta_{12},\Delta_{34}}_{\Delta,J}(z,\bar{z})
=\big(\Delta(\Delta-d)+J(J+d-2)\big)G^{\Delta_{12},\Delta_{34}}_{\Delta,J}(z,\bar{z})\,,\label{Casi}
\eea
where (denoting $a=-\fft12\Delta_{12}$, $b=\fft12\Delta_{34}$)
\bea
&& \mathcal{C}_2=\mathcal{D}_z+\mathcal{D}_{\bar{z}}+2(d-2)\fft{z\bar{z}}{z-\bar{z}}
((1-z)\partial_z-(1-\bar{z})\partial_{\bar{z}})\,,
\cr && \mathcal{D}_z=2(z^2(1-z)\partial_z^2-(1+a+b)z^2\partial_z-abz)\,.\label{D}
\eea
In even $d$ dimensions, the exact solutions of Casimir equation (\ref{Casi}) can be obtained, e.g.~in $d=4$, it is
\bea
&& G^{\Delta_{12},\Delta_{34}}_{\Delta,J}=\fft{z\bar{z}}{z-\bar{z}}(k_{\Delta+J}(z)k_{\Delta-J-2}(\bar{z})
-k_{\Delta+J}(\bar{z})k_{\Delta-J-2}(z))\,,
\eea
where $k_{\beta}(z)$ is hypergeometric function
\be
 k_\beta(z)=z^{\fft{\beta}{2}}\,_2F_1[\ft12{\beta}+a,\ft12{\beta}+b;\beta;z]\,.
\ee

In this paper, our focus is on $a=b=0$. In general dimensions, although the exact solutions are hard to come by, various series expansion can be applied to reveal the information encoded in the conformal blocks. In this paper, we consider the OPE limit where $z\ll1$ and $\bar{z}\ll1$ (note $(z,\bar{z})$ is with respect to $t$-channel version (\ref{t-channel2})), for which the leading OPE tells us the conformal blocks can take the simple form
\be
G^{00}_{\Delta,J}= (z\bar{z})^{\fft{\Delta}{2}}\fft{J!}{(\fft{d}{2}-1)_J}C^{\fft{d}{2}-1}_J
(\fft{z+\bar{z}}{2\sqrt{z\bar{z}}})+\cdots\,,\label{G00}
\ee
where $C^{\fft{d}{2}-1}_J$ is the Gegenbauer polynomial and it can be expressed in terms of hypergeometric function
\be
C_J^{\fft{d}{2}-1}(x)=\fft{(d-2)_J}{J!}\,_2F_1[-J,d+J-2;\ft12(d-1);\ft12(1-x)]\,.\label{C-F}
\ee
In fact for an integer $J$, the hypergeometric function reduces to some $x$ polynomials of finite order.
Note (\ref{G00}) is the main ingredient we use in this paper to compare with the bulk calculation and read off the OPE coefficients.

For the truncation analysis this paper, as was discussed in detail in sections \ref{OPE} and \ref{OPEcharge}, it is more convenient to make further approximation. Imposing the light-cone limit where $z\ll\bar{z}\ll1$, (\ref{G00}) together with (\ref{C-F}) gives us the power law
\be
G^{00}_{\Delta,J}=z^{\fft{\Delta-J}{2}}\bar{z}^{\fft{\Delta+J}{2}}(1+\mathcal{O}(\bar{z})+\mathcal{O}
(\fft{z}{\bar{z}})+\cdots)\,.\label{powerlaw}
\ee

\section{Scalar equation}
\label{freescalar}

In this appendix, we present the scalar equation (\ref{phi1}) and more general equation (\ref{phi3}) with the coordinate variables (\ref{change-va}). In the background (\ref{metric}), explicitly, we have
\bea
&& \Big(\frac{1}{r^2f}\partial_t^2+r^2 h \big(\frac{f'}{2f}+\frac{h'}{2h}+\frac{5}{r}+\partial_r\big)\partial_r + \frac{1}{r^2}\big(\frac{2}{u}+\partial_u\big)\partial_u-m^2
\cr &&
\cr
&&
+(d-4)\big(\frac{1}{ur^2} \partial_u+h r \partial_r \big)\Big)\Phi(r,t,u)=0\,,
\eea
where for (\ref{phi1}) we have $m^2=\Delta_L(\Delta_L-d)$, while for (\ref{phi3}) $m^2$ is a function of $r$ and the explicit form depends on theory detail.
Changing the variables to (\ref{change-va}) and factorizing the AdS propagator as in (\ref{factorize}), we can present the equation for $G$ explicitly as follows
\bea
&&\partial_r^2 G+\frac{w^2-\hat{u}^2-1+(\hat{u}^2+(w^2-1)^2h)f}{r^2w^2fh}\partial^2_w G+\frac{1+\hat{u}^2 h}{r^2h}\partial^2_{\hat{u}}G\cr
&& +\frac{2\hat{u}(1+(w^2-1)h)}{r^2 w h}\partial_w\partial_{\hat{u}}G +\frac{2\hat{u}}{r}\partial_r \partial_{\hat{u}}G+\frac{2(w^2-1)}{r w}\partial_r \partial_w G\cr
&&+\Big(\frac{(fh)'}{2fh}+\frac{w^2(1+d-2\Delta_L)+4\Delta_L}{r w^2}\Big)\partial_r G
\cr
&&+\fft{\partial_w G}{r^2 w^3 fh} \Big(\ft12 w^2 (w^2-1) r (hf)' + 1 + \hat u^2 + \Delta(4-4w^2 + 4 \hat u^2)
+(d-1) w^2 f\cr
&&-(4\Delta_L+1) \hat u^2 f + (w^2-1)(1 + w^2(d+1-2\Delta_L) + 4\Delta_L) h f
\Big)
\cr
&&+\fft{\partial_{\hat{u}}G}{r^2w^2 \hat u f h}\Big(\ft12 r^2 w^2 (hf)' + \big((d-2) w^2 - 4 \Delta_L \hat u^2\big) f +
(w^2 (d+1-2\Delta_L) + 4 \Delta_L) \hat u^2 f h \Big)\cr
&& +\fft{\Delta_L}{r^2 w^4 hf}\Big(\ft12 w^2(2-w^2) r (hf)' -2 \left((2 \Delta_L +1) w^2-2 (\Delta_L +1) \left(\hat u ^2+1\right)\right)(f-1)\cr
&&+\left(w^4 (\Delta_L -d)+2 w^2 (d-2- 2\Delta_L)+4 (\Delta_L +1)\right) (h-1) f
\Big)G=0\,.\label{eqG}
\eea
It is clear that for the AdS vacuum with $f=1=h$, the last term vanishes and we have constant $G$.  For general $h$ and $f$, the equation can be solved by first making a decomposition (\ref{ansatz1}). Then for non-integer $\Delta_L$, $G^L(r,w,\hat u)$ independently satisfies precisely the linear equation (\ref{eqG}) and hence the coefficients in the polynomial modes expansion cannot be fully determined by the background metric functions. The $G^T(r,w,\hat u)$ equation, on the other hand, becomes nonhomogeneous with a source, namely
\bea
&&\partial_r^2 G^T+\frac{w^2-\hat{u}^2-1+(\hat{u}^2+(w^2-1)^2h)f}{r^2w^2fh}\partial^2_w G^T+\frac{1+\hat{u}^2 h}{r^2h}\partial^2_{\hat{u}}G^T\cr
&& +\frac{2\hat{u}(1+(w^2-1)h)}{r^2 w h}\partial_w\partial_{\hat{u}}G^T +\frac{2\hat{u}}{r}\partial_r \partial_{\hat{u}}G^T+\frac{2(w^2-1)}{r w}\partial_r \partial_w G^T\cr
&&+\Big(\frac{(fh)'}{2fh}+\frac{w^2(1+d-2\Delta_L)+4\Delta_L}{r w^2}\Big)\partial_r G^T
\cr
&&+\fft{\partial_w G^T}{r^2 w^3 fh} \Big(\ft12 w^2 (w^2-1) r (hf)' + 1 + \hat u^2 + \Delta(4-4w^2 + 4 \hat u^2)
+(d-1) w^2 f\cr
&&-(4\Delta_L+1) \hat u^2 f + (w^2-1)(1 + w^2(d+1-2\Delta_L) + 4\Delta_L) h f
\Big)
\cr
&&+\fft{\partial_{\hat{u}}G^T}{r^2w^2 \hat u f h}\Big(\ft12 r^2 w^2 (hf)' + \big((d-2) w^2 - 4 \Delta \hat u^2\big) f +
(w^2 (d+1-2\Delta) + 4 \Delta) \hat u^2 f h \Big)\cr
&& +\fft{\Delta_L}{r^2 w^4 hf}\Big(\ft12 w^2(2-w^2) r (hf)' -2 \left((2 \Delta_L +1) w^2-2 (\Delta_L +1) \left(\hat u ^2+1\right)\right)(f-1)\cr
&&+\left(w^4 (\Delta_L -d)+2 w^2 (d-2- 2\Delta_L)+4 (\Delta_L +1)\right) (h-1) f
\Big)G^T=\nn\\
&& -\fft{\Delta_L}{r^2 w^4 hf}\Big(\ft12 w^2(2-w^2) r (hf)' -2 \left((2 \Delta_L +1) w^2-2 (\Delta_L +1) \left(\hat u ^2+1\right)\right)(f-1)\cr
&&+\left(w^4 (\Delta_L -d)+2 w^2 (d-2- 2\Delta_L)+4 (\Delta_L +1)\right) (h-1) f
\Big)\,.\label{eqGT}
\eea
This implies that we can determine $G^T(r,w,\hat u)$ order by order in terms of the source in the right-hand-side of the equation that depends only on the metric functions $(h,f)$.  In other words, if we write (\ref{eqG}) as some linear differential operator acting on $G$, namely $\hat L * G=0$, then the decomposition (\ref{ansatz1}) for non-integer $\Delta_L$ implies
\be
\hat L * G^L=0\,,\qquad \hat L * G^T = - \hat L * 1\,.
\ee
The second equation contains a source that determines $G^T$.

The situation becomes more complicated when $\Delta_L$ is an integer. The analysis of the scalar equation in this case is given in section \ref{integerDelta}.

\end{document}